\documentclass[letter,openany]{aa}  
\usepackage{graphicx}
\usepackage{grffile}
\usepackage{soul}
\usepackage{newtxtext}
\usepackage{amsbsy}
\usepackage{wasysym}
\usepackage{natbib}
\usepackage{mathtools}
\usepackage{amsmath}
\usepackage{longtable}
\usepackage{pdflscape}
\usepackage{afterpage}
\usepackage[colorlinks=True,allcolors=cyan]{hyperref}
\bibpunct{(}{)}{;}{a}{}{,}
\usepackage{subcaption}
\usepackage{caption}
\usepackage{multicol}
\usepackage{wrapfig}
\usepackage{stfloats}

\newcommand{\kms}{km\,s$^{-1}$}

\newcommand{\gaia}{{\it Gaia} }

\begin{document} 

\title{The distance to the Serpens South cluster from H$_2$O masers}

   \author{Gisela \ N.\ Ortiz-Le\'on \inst{1,2}
           \and
           Sergio A.\,Dzib\inst{3, 2} 
           \and
           Laurent Loinard\inst{4,5} 
           \and
          Yan Gong\inst{2}
          \and
          Thushara Pillai\inst{6}
          \and
          Adele Plunkett\inst{7}
          }
          %

   \institute{Instituto Nacional de Astrof\'isica, \'Optica y Electr\'onica, Apartado Postal 51 y 216, 72000, Puebla, Mexico.
              \email{gortiz@inaoep.mx}
              \and
               Max Planck Institut f\"ur Radioastronomie, Auf dem H\"ugel 69,  D-53121 Bonn, Germany.
               \and
            Institut de radioastronomie millim\'etrique, 300 rue de la piscine, 38406 Saint Martin d'H\`eres, France.
            \and 
            Instituto de Radioastronom\'ia y Astrof\'isica, Universidad Nacional Aut\'onoma de M\'exico, Morelia, 58089, Mexico.
            \and
            Instituto de Astronom{\'\i}a, Universidad Nacional Aut\'onoma de M\'exico, Apdo Postal 70-264, Ciudad de M\'exico, Mexico.
              \and
            MIT Haystack Observatory, 99 Millstone Road, Westford, MA 01886, USA
            \and
            National Radio Astronomy Observatory (NRAO), 520 Edgemont Road, Charlottesville, VA 22903, USA
               }


\titlerunning{Distance Serpens South } 
\abstract{
In this Letter we report Very Long Baseline Array observations of 22~GHz water masers toward the protostar \object{CARMA--6}, located at the center of the Serpens South young cluster. From the astrometric fits to maser spots, we derive a distance of $440.7\pm$3.5~pc for the protostar ($1\%$ error). This represents the best direct distance determination obtained so far for an object this young and deeply embedded in this highly obscured region. Taking depth effects into account, we obtain a distance to the cluster of $440.7\pm4.6$~pc.
Stars visible in the optical that have astrometric solutions in the \gaia Data Release~3 are, on the other hand, all located at the periphery of the cluster. Their mean distance of $437^{+51}_{-41}$~pc is consistent within $1\sigma$ with the value derived from maser astrometry.    
As the maser source is at the center of Serpens South, we finally solve the ambiguity of the distance to this region that has prevailed over the years.
}

\maketitle

\section{Introduction}\label{sec:intro}

The \object{Serpens South region} hosts a rich protostellar cluster that was discovered by \cite{Gutermuth2008} from {\it Spitzer} imaging. 
The cluster is embedded in a massive molecular filament \citep{Andre2010,Tanaka2013} and contains approximately 400 young members (see the recent work by \citealt{Sun2022}), of which very few are visible in the optical \citep{Herczeg2019}.
Since its discovery, this protocluster has become a favorite target for observations of young stars in the earliest phases of their development \citep[e.g.,][]{Maury2011} and studies of related phenomena, such as molecular gas kinematics \citep[e.g.,][]{Fernandez-Lopez2014,Friesen2016,Dhabal2018}, protostellar outflows \citep{Plunkett2015ApJ}, and magnetic fields \citep{Pillai2020}. Serpens South is the most active region, in terms of star formation, projected in the direction of a larger complex of molecular clouds  known as the Aquila Rift (or the Serpens-Aquila Rift). In the vicinity of Serpens South ($\sim20'$ to the east) resides the \object{W40} region \citep{Westerhout1958}, which hosts an HII region, hundreds of candidate young stars \cite[e.g.,][]{Povich2013,Sun2022}, and several OB stars \citep{Shuping2012}. The well-studied Serpens Main molecular cloud is located about 3 degrees north of Serpens South and is also an active star-forming region in the Serpens-Aquila Rift \cite[see the review by][]{Eiroa2008}.

The distance to the \object{Serpens South cluster} has been debated over the years. Early works assumed a distance of 260 pc, the estimated distance to the front clouds in the line of sight toward the Aquila Rift (\citealt{Straizys2003}, see also the discussion by \citealt{dzib2010}). 
However, later measurements of the distance to star-forming regions within the Aquila Rift, including Serpens South, are in the range $\approx380-480$~pc \citep[][]{dzib2010,Ortiz2017,Herczeg2019,Zucker2019}. For instance, \cite{Herczeg2019} reported distances to subclusters and dark clouds in the Aquila Rift based on \gaia Data Release 2 (DR2) data and star counts. These authors found distances of 438$\pm$11, 478$\pm$6, 383$\pm$2, and 407$\pm$16~pc for Serpens Main, Serpens Northeast, Serpens Far-South, and LDN~673, respectively. In addition, they reported larger distances of $\sim$500-700~pc for the dark clouds (which they refer to as the Aquila Rift) located  northeast and southwest of the active star-forming regions. 

More recently, \cite{Anderson2022} used \gaia Early Data Release 3 (EDR3) data from the Serpens South and W40 regions and arrived at a mean distance of $455\pm50$~pc for this region. We note that this is currently the most direct estimation of the distance to the Serpens South cluster, as it is directly derived from the parallaxes. However, the error bar is large and reflects the dispersion on the parallax measurements, as these regions are highly obscured at the \gaia bands. On the other hand, \cite{Ortiz2018} measured Very Long Baseline Array (VLBA) parallaxes of eight radio continuum sources in the star-forming regions Serpens Main and W40, both within the Aquila Rift, resulting in a mean distance of 436.0$\pm$9.2~pc. 

It is important to note that, using the VLBA, \cite{Ortiz2017} and \cite{Ortiz2018} did not detect any radio continuum sources associated with the Serpens South region. However, based on \gaia DR2 parallaxes to two stars in the outskirts of the Serpens South region and their similar values to the VLBA parallaxes of Serpens Main and W40 stars, \cite{Ortiz2018} conclude that the three regions reside at the same distance. On the other hand, estimations based on studies of the X-ray luminosity function of Serpens South have placed this region at the significantly closer distance of 260 pc \citep{winston2018}. Following their indirect distance estimation, \citet{winston2018} suggest that Serpens South is at the front edge of the Aquila Rift complex, closer than W40 and Serpens Main and not physically associated with them. 
Other authors have also assumed a closer distance (e.g., $250$~pc by \citealt{Maury2019} and $350$~pc by \citealt{galametz2019} and \citealt{podio2021}). 
As there are no direct distance determinations to the youngest stars in this region, these results did not resolve the debate regarding the distance to the Serpens South cluster. However, using incorrect distances affects the estimates of envelope and disk masses, as well as of the outflow properties of their stellar members.
For instance, using a distance of 436~pc would increase internal luminosities and envelope masses reported in \cite{Maury2011} by a factor of 3.

Recently, \cite{Ortiz2021AJ}  discovered 22 GHz water maser emission from two protostars in the Serpens South cluster, CARMA--6 and \object{CARMA--7}. 
These masers offer an opportunity to attempt an independent and direct measurement of the distance to Serpens South.
In this Letter we present the results from VLBA follow-up observations of the 22~GHz water masers associated with CARMA--6, which we use to derive the distance to the Serpens South cluster. Section \ref{sec:obs} describes the observations and data calibration. In Sect. \ref{sec:results} we present the astrometric fits. Finally, our results are discussed in Sect. \ref{sec:discussion}.

\section{Observations and data reduction}\label{sec:obs}

We observed  CARMA--6 with the VLBA during a total of 12 epochs between September 2020 and September 2021 (Table \ref{tab:obs-vlba}). Two epochs were reported in \cite{Ortiz2021AJ}, while the rest are analyzed here for the first time. 
The phase center was at position $\alpha$(J2000) = 18:30:03.538, $\delta$(J2000) = --02:03:08.377.  The data were taken at 22.2 GHz with four intermediate frequency bands of 16~MHz bandwidth each. The third band was centered at the $6_{1,6} - 5_{2,3}$ H$_2$O transition  (rest frequency 22,235.080 MHz) and correlated with a spectral resolution of 15.625~kHz, corresponding to a resolution of 
$\sim$0.2~km~s$^{-1}$ in velocity.  

We performed fast-switching observations between our target and the phase reference calibrator \object{J1824+0119} ($\alpha$(J2000) = 18:24:48.143436, $\delta$(J2000) = +01:19:34.20183),  switching sources every $\approx$30 seconds.  Additional 30-minute blocks of calibrators distributed over a wide range of elevations were observed at 23.7~GHz every $\approx$2 hours.

The data were calibrated with the Astronomical Imaging System (AIPS; \citealt{Greisen2003}), using the ParselTongue scripting interface \citep{Kettenis2006} and following standard procedures for phase-referencing observations \citep[e.g.,][]{Reid2009}. 
As the masers were variable in both flux density and velocity (see Sect. \ref{sec:spots}), 
we opted to derive the fringe-fitting solutions from the scans on the phase-reference calibrator and then applied the solutions to the target. The calibrated data were imaged in individual channels using $4096\times4096$ pixels and a pixel size of $50~\mu$as. The sensitivities and beam sizes of the resulting data cubes are given in Table \ref{tab:obs-vlba}.
Spot positions and fluxes were determined by fitting a Gaussian to the brightness distribution at individual channels using the AIPS task {\tt jmfit}. We note that the expected statistical positional errors are on the order of 70~$\mu$as for emission detected at S/N=5.

\section{Results}\label{sec:results}

\subsection{Maser spots}\label{sec:spots}

\begin{figure}[!bht]
\begin{center}
 {\includegraphics[width=0.2\textwidth,angle=0]{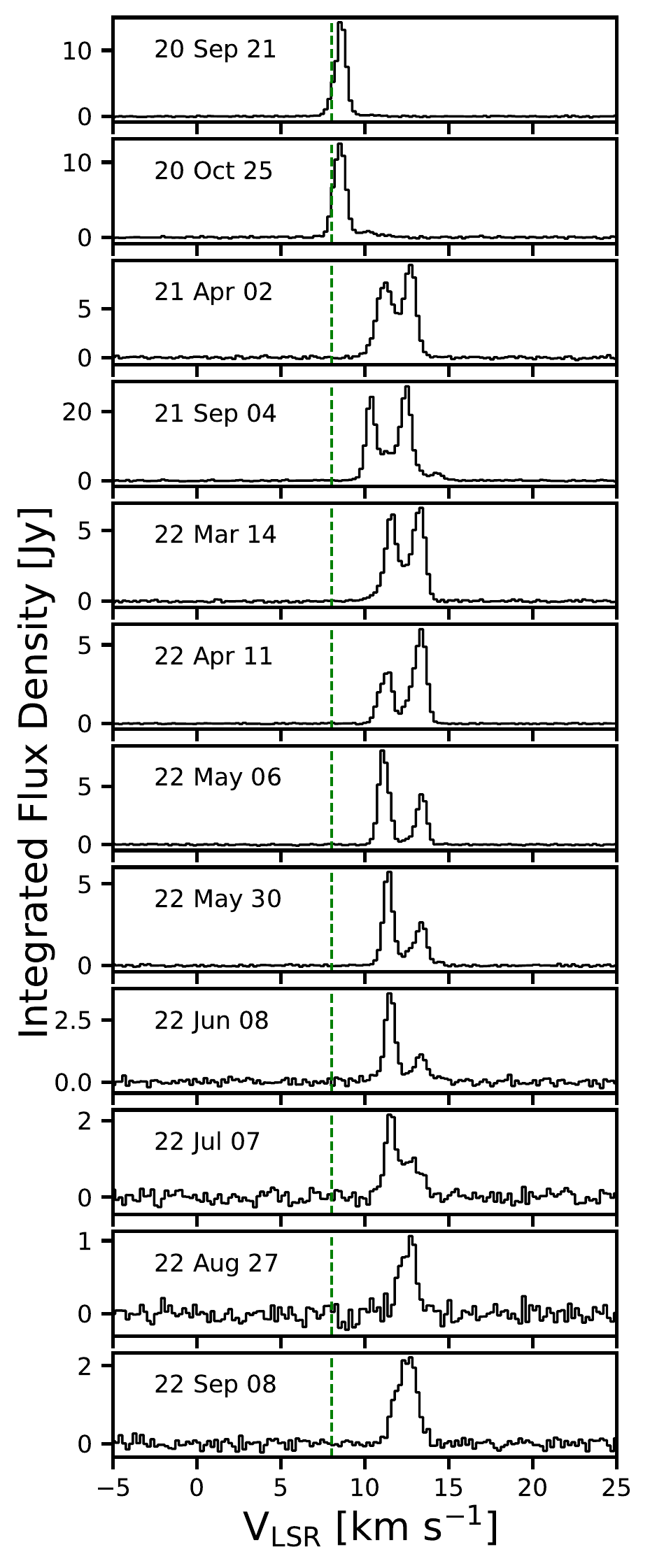}}
 \end{center}
\caption{Averaged VLBA spectra of the water maser emission observed toward CARMA--6. They were obtained by integrating over the area that covers all detected spots, and their epochs are indicated in the legends. The vertical dashed line at $V_{\rm LSR}$=8~\kms ~marks the systemic velocity of the cloud. }
\label{fig:spectra}
\end{figure}

Spectra  of water emission from CARMA--6 are shown in Fig. \ref{fig:spectra}, where the $x$-axis represents the radial velocity  with respect to the local standard of rest (LSR), $V_{\rm LSR}$. The spectra display three main velocity components, at $V_{\rm LSR} \approx 8.5$, 11, and 13~\kms, ~that change over time and are redshifted with respect to the systemic velocity of the cloud, $V_{\rm sys} \approx 8$~\kms ~\citep{Kirk2013,Plunkett2015ApJ}. The 8.5~\kms ~component disappeared after our observation in October 2020, while the other two components became brighter than before. Overall, the intensity of the emission starts declining after the maximum reached in September 2021. 

Figures \ref{fig:spots} and \ref{fig:spots-2} show the distribution of the spot positions measured with {\tt jmfit} at individual velocity channels during all observed epochs. 
At velocities from $V_{\rm LSR} =9.76$ to 10.81~\kms, ~we clearly see that the spots describe a helical motion due to the combined effect of the parallax and proper motion. From $V_{\rm LSR} =11.02$ to 13.34~\kms~the detection of many spots per epoch hampers the identification of motion of individual spots.  It is not clear why many more maser spots (and, correspondingly, the highest maser intensity) appear at these velocities, but this suggests that the water masers could be excited in 
internal shocks of the redshifted flow emerging from the protostar. Outside the $9.76-13.97$~\kms velocity range, the spots persist over fewer epochs due to variability.

\subsection{Astrometric fits}\label{sec:astrometry}

\begin{table*}
\tiny
\caption{Astrometric fits.}
\label{tab:fits} 
\centering 
\begin{tabular}{c c c c c c c c }  
\hline\hline  
Velocity channels & Parallax & $\mu_\alpha \cos\delta$ & $\mu_\delta$ & RA\ rms & Decl.\ rms & Distance \\
(\kms) & (mas) & (mas~yr$^{-1}$) & (mas~yr$^{-1}$) &(mas) &(mas) & (pc) \\
\hline 
9.76 -- 10.81$^{a}$  & $2.280 \pm 0.233$ & $-2.815 \pm 0.334$ & $-6.788 \pm 0.177$ & 0.72 & 0.35  & $439\pm45$ \\ 
9.76 -- 10.81$^{b}$  & $2.287 \pm 0.072$ & $-1.324 \pm 0.085$ & $-7.382 \pm 0.151$ & 0.11 & 0.21  & $437\pm14$ \\
10.40   & $2.275 \pm 0.038$ & $-1.204 \pm 0.039$ & $-7.447 \pm 0.128$ & 0.010 & 0.040 & $439.6\pm7.4$ \\ 
10.60   & $2.265 \pm 0.021$ & $-1.063 \pm 0.025$ & $-7.744 \pm 0.075$ & 0.006 & 0.019 & $441.5\pm4.1$ \\
10.81   & $2.332 \pm 0.191$ & $-1.321 \pm 0.314$ & $-7.580 \pm 0.273$ & 0.077 & 0.076 & $429\pm35$ \\ 
\hline 
\end{tabular}
\tablefoot{
\tablefoottext{a}{Fit to maser spots detected in epochs 1 to 9.}
\tablefoottext{b}{Fit to maser spots detected in epochs 4 to 9.}
}
\end{table*}

\begin{figure*}[!bht]
\begin{center}
 {\includegraphics[width=0.8\textwidth,angle=0]{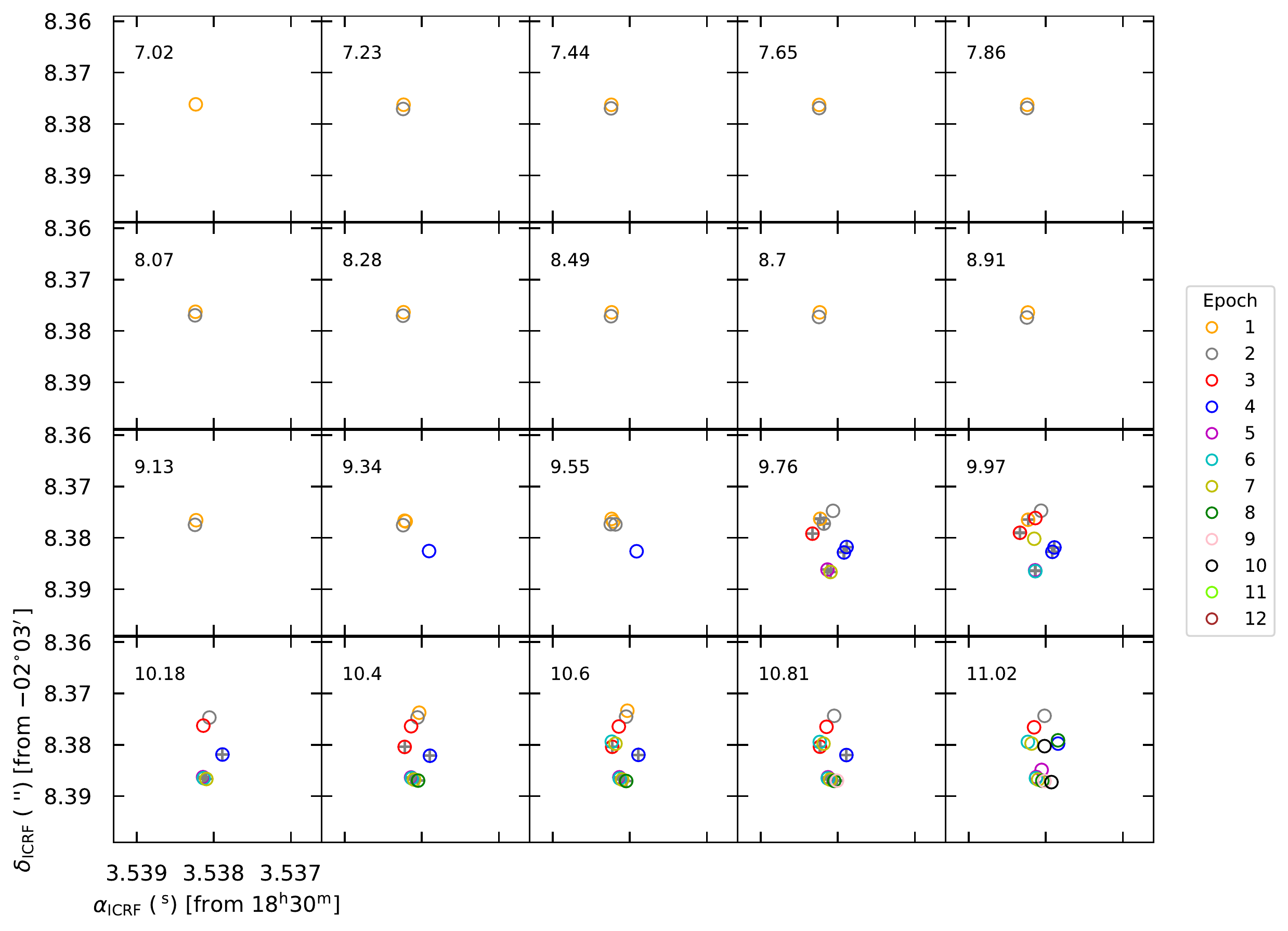}}
\caption{Spatial distribution of the water maser spots. The numbers in the corners indicate the velocity channel in~\kms. 
The symbols are color-coded by epoch (see the figure legend). The spots used for the astrometric fits are filled with crosses (see Sect. \ref{sec:astrometry}). 
}
\label{fig:spots}
\end{center}
\end{figure*}

To perform the astrometric fittings, following \cite{Sanna2017}, we first selected maser spots that persist for five epochs or more and that can be distinguished as isolated maser centers. We note that we did not attempt to fit spots detected at velocities from  $V_{\rm LSR} =11.02$ to 13.34~\kms, because at these velocities we detect many spots at multiple positions in the same epoch (see Fig. \ref{fig:spots}), which hampers the spot identification.
The spots that satisfy the criteria given above and that are outside the 11.02 to 13.34~\kms ~range are those at  $V_{\rm LSR}$ = 9.76, 9.97, 10.18, 10.4, 10.6, and 10.81~\kms. The positions of these spots are given in Table \ref{tab:jmfit}.

Considering only the velocity range $V_{\rm LSR} =9.76$ to 10.81~\kms, we see in Fig. \ref{fig:spots} that for epochs 5 to 9  the group of spots detected toward the south are tracing the same gas cloudlet. These spots have similar positions, as expected given that the observations were taken $\sim$1~month apart.  Thus, we discarded the spots detected toward the north ($\delta \approx$ --02:03:08.380) in epochs 6 and 7. On the other hand, for epochs 1, 2, and 3 we used the spots that are closer in position to the spots detected in the rest of the epochs, thus discarding those detected at $\delta \approx$ --02:03:08.375 (spots that are discarded are plotted with empty symbols in Fig. \ref{fig:spots}, while those used for the astrometric fits are filled with a cross).
Using the least-squares method \citep[e.g.,][]{Loinard2007}, we fit the positions of the remaining spots with a model that accounts for the effect of parallax ($\varpi$) and proper motions ($\mu_\alpha, \mu_\delta$). The position uncertainties used in the fitting procedure are inflated with error floors that yield a reduced $\chi^2$ of near unity  (e.g., \citealt{Ortiz2017,Sanna2017}). Then, we used the covariance matrix to obtain the errors on the astrometric parameters. Table \ref{tab:fits} gives the best-fit astrometric parameters and their associated uncertainties. 

\begin{figure*}[tbh]
  \begin{subfigure}[t]{0.24\textwidth}
  \centering
  \includegraphics[scale=0.32,angle=0]{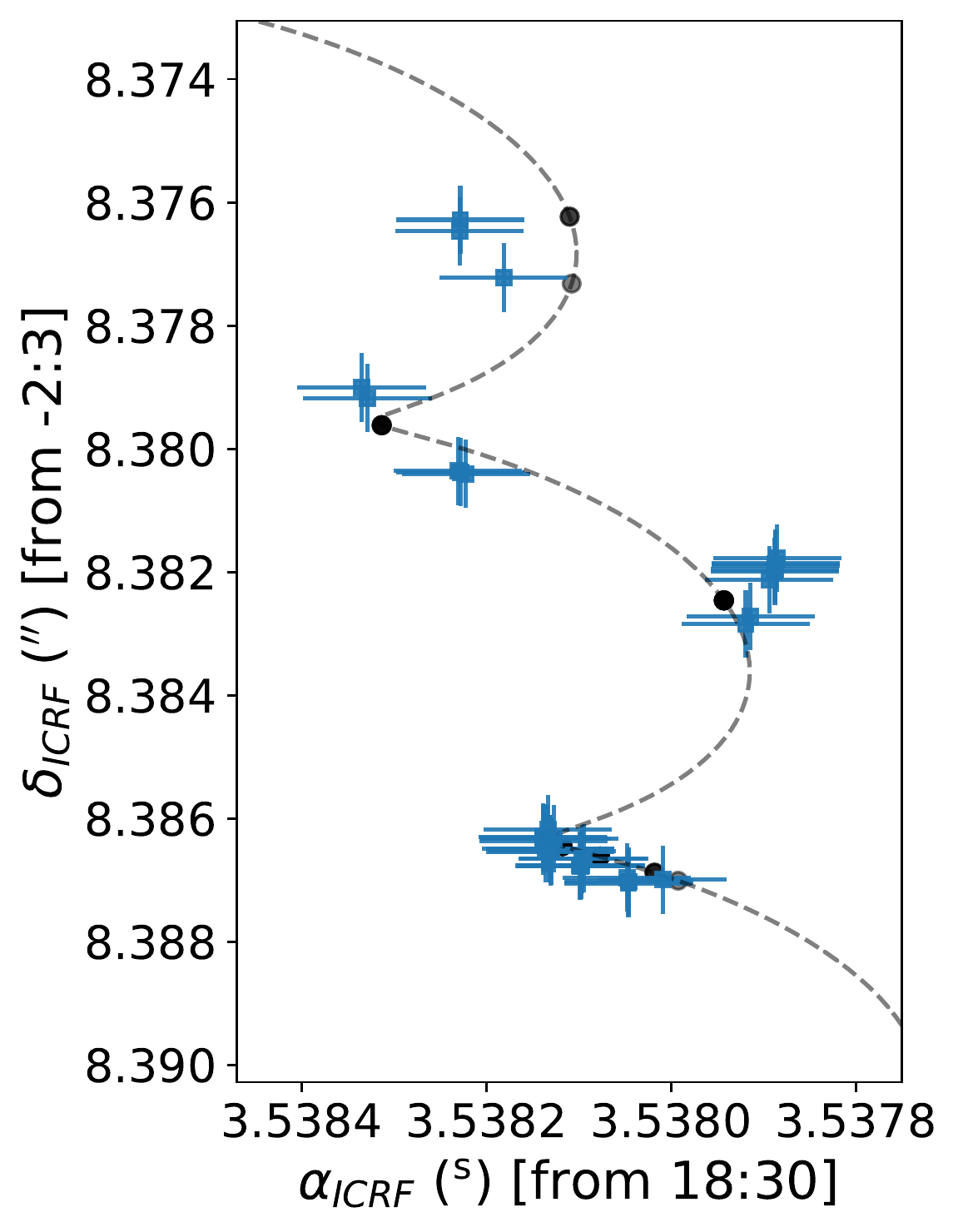}
  \caption{}
  \end{subfigure}
  \begin{subfigure}[t]{0.36\textwidth}
  \centering
  \includegraphics[scale=0.32,angle=0]{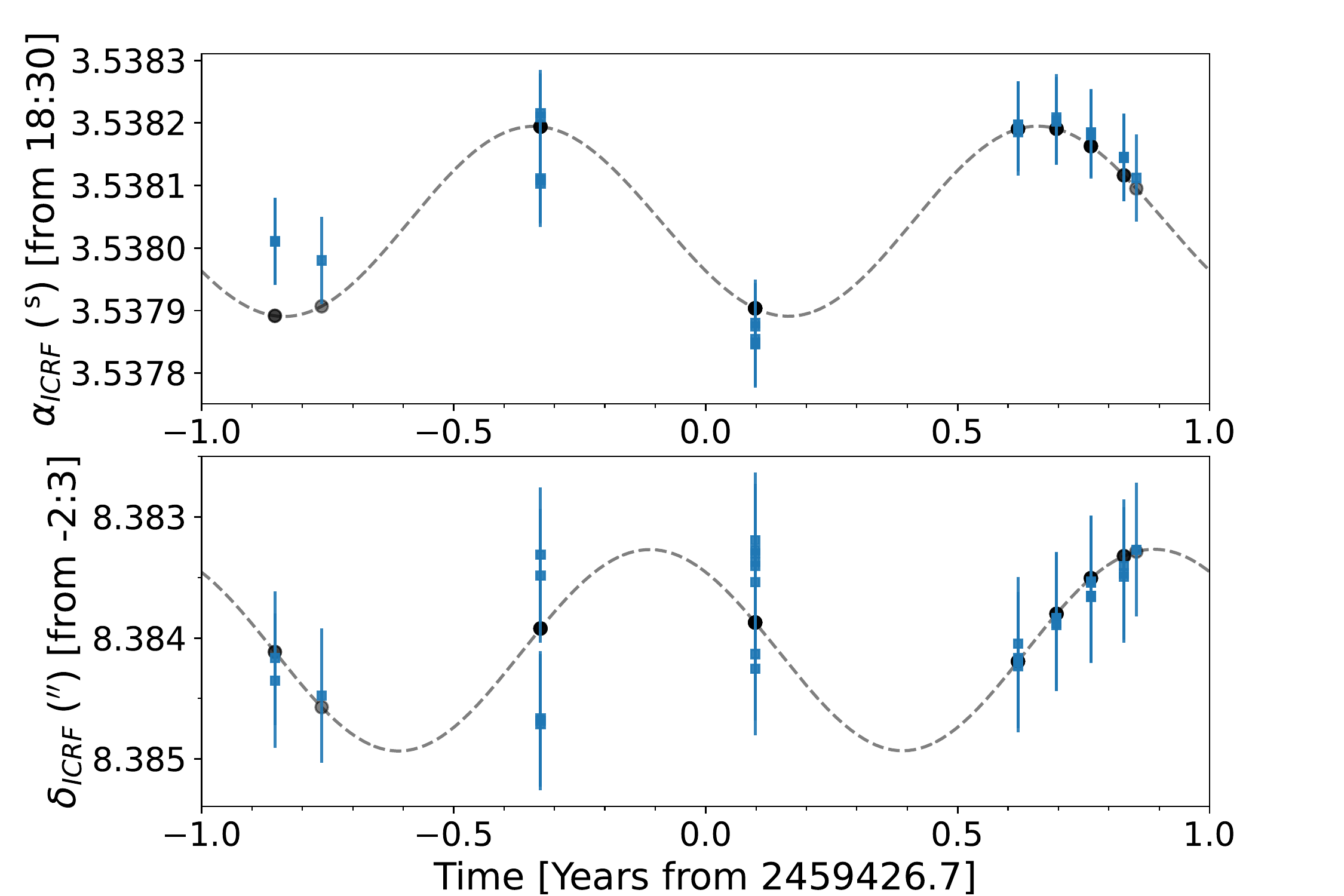}
  \caption{}
  \end{subfigure} 
  \begin{subfigure}[t]{0.3\textwidth} 
  \centering
  \includegraphics[scale=0.32,angle=0]{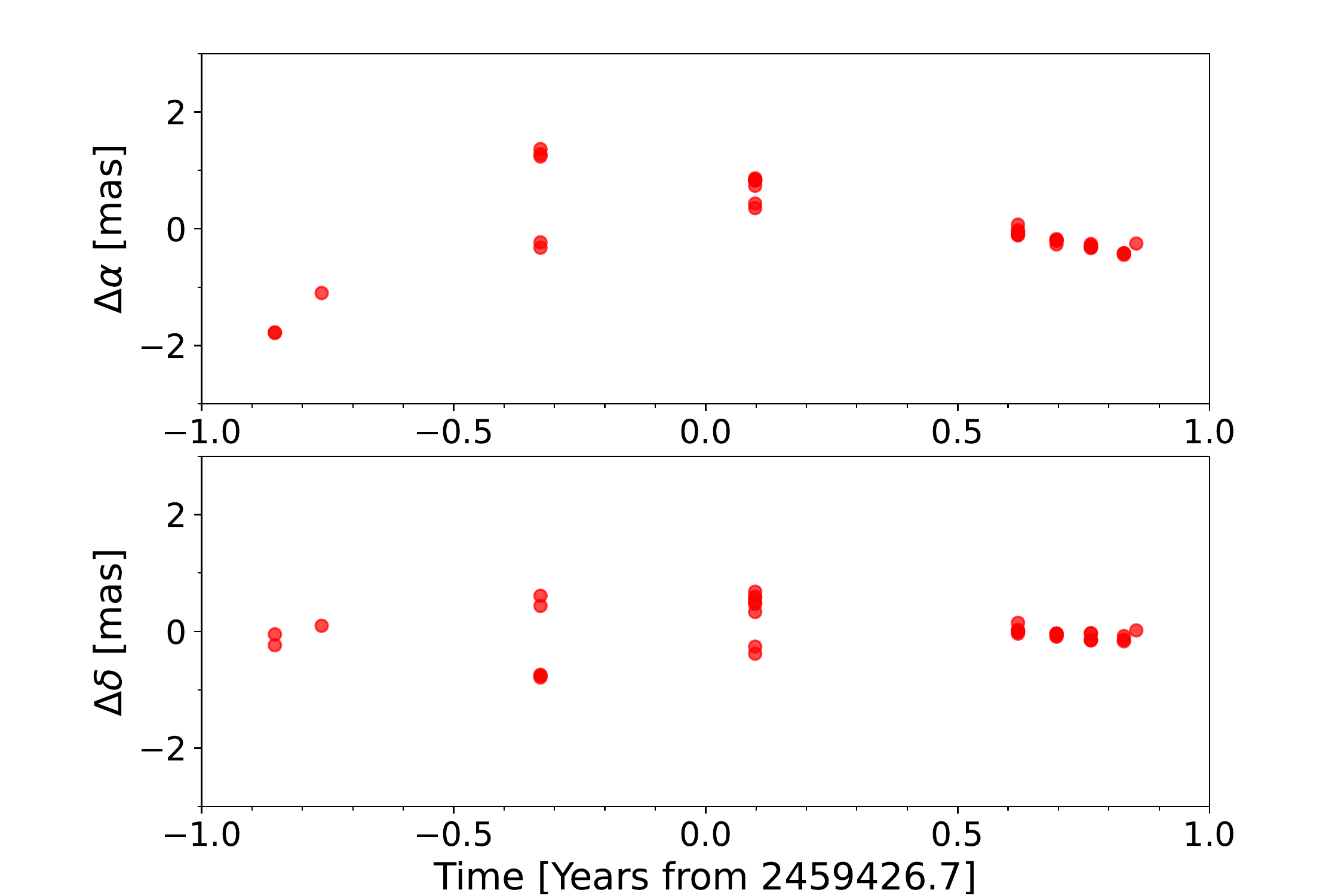}
  \caption{}
  \end{subfigure} 
\caption{Astrometric fits to  water maser spots detected in the velocity range $V_{\rm LSR} =9.76$ to 10.81~\kms. (a) Measured absolute spot positions, shown as blue squares. The dashed black line is the fitted model. Expected positions from the best-fit model are shown as black dots. (b) Measured positions and best-fit model after proper motions are removed. (c) Residuals of the astrometric fit in RA\  (top panel) and Decl.\ (bottom panel). }
\label{fig:fit-combined}
\end{figure*}

The astrometric fitting to the set of spots selected as described above gives a best-fit parallax of $2.263 \pm 0.231$~mas. Figure \ref{fig:fit-combined} shows the best-fit model and the measured positions of the spots in the plane of the sky (panel a). We also show the positions of the spots as a function of time after removing proper motions (panel b) and the residuals from the fit (panel c). The residuals are large, with post-fit rms values of 0.71 and 0.33~mas in RA and Decl., respectively, which suggests that we are fitting spots that do not trace the same gas cloudlet. It is clear that the spots that show the largest residuals, mainly in the RA\ direction, are those detected in the first three epochs. Discarding these spots and fitting only those detected in epochs 4 to 9 gives a parallax of $2.276 \pm 0.072$~mas (Fig. \ref{fig:fit-combined-2}). The residuals are smaller than those obtained in the previous fit, with post-fit rms values of 0.11 and 0.21~mas in RA and Decl., respectively, while the parallax uncertainty decreases by a factor of $\approx$3. The rms values are still high in declination.

Since the spots are not necessarily at the same position in the whole 9.76--10.81~\kms ~velocity range, we fit spots at individual velocity channels when more than five points were available. Table \ref{tab:fits} gives the best-fit astrometric parameters for velocity channels at 10.4, 10.6, and 10.81~\kms. These fits are shown in Figs.~\ref{fig:fit-104}, \ref{fig:fit-106}, and \ref{fig:fit-1081}. We find that the velocity channels at 10.4 and 10.6~\kms ~have the smallest post-fit rms values. Thus, they yield a better fit to the data. We also note that all fits yield similar solutions for the parallax but different values for the proper motion, confirming that the spots in individual velocity channels are most likely tracing different gas regions.

In Fig. \ref{fig:spots-woprlx} we show the distribution of maser spots after removing the parallax from the fit to the 9.76--10.81~\kms ~velocity range. 
The dashed red line in this figure shows the best-fit model with the parallax signature removed, that is to say, it reflects only the proper motions (see Table \ref{tab:fits}).
The spots in the velocity range $V_{\rm LSR}$ = 9.76 to 10.81~\kms ~closely follow this line, that is, they have a linear motion toward the southwest 
that results from the effect of a uniform proper motion. 
We also see that most spots in the velocity range $V_{\rm LSR}$ = 11.23 to 15.02~\kms ~do not share this motion. Indeed, spots in the 12.5--14.18~\kms ~range are moving mostly to the south. We can conclude that these spots are tracing distinct gas cloudlets.  

In Appendix \ref{sec:companion} 
we consider the possibility of an unseen companion
in the fits to the data from the nine epochs and the full 9.76 to 10.81~\kms velocity range.  When this is compared with the fits presented above, we find that additional terms in the astrometric model do not improve the fits (Fig.~\ref{fig:fit-acceleration}). This suggests that the large residuals in Fig.~\ref{fig:fit-combined} are not due to orbital motion, but due to combining positions from different velocity channels. 

\section{Discussion}\label{sec:discussion}

\subsection{The distance to Serpens South}

Inverting the weighted mean average of the parallaxes reported in Table~\ref{tab:fits} gives a distance of $440.7\pm$3.5~pc,
where the uncertainty is the standard error of the weighted mean.  
This value is consistent within 1$\sigma$ with the distance reported by \cite{Anderson2022}, who used \gaia EDR3 parallaxes in Serpens South and W40.
However, the new distance is an improvement (by an order of magnitude) over the \gaia EDR3 measurement. In Appendix \ref{sec:gaia} we collect parallaxes of young stellar objects (YSOs) from the most recent data release, DR3. Although the dispersion of these parallaxes is large (Figs. \ref{fig:hist-gaia} and \ref{fig:hist-gaia-2}), we find that their weighted mean is  consistent with the maser parallax.  

From parallax measurements to eight radio continuum sources in W40 and Serpens Main 
, \cite{Ortiz2018} obtained a mean distance of $436.0\pm9.2$~pc and individual distances consistent with one another within the errors. 
Our measurement is consistent with values found previously by \cite{Ortiz2018} using VLBA parallaxes of radio continuum sources in Serpens Main and W40. 
Thus, the parallax to one star embedded in the Serpens South cluster obtained in this work suggests that Serpens South is at the same distance as Serpens Main and W40, confirming that the three structures are physically associated. 

\subsection{Proper motions}

Figure \ref{fig:gaia-appendix} shows the proper motions of the maser source and the \gaia sources.  The \gaia sources have a mean value of $(\overline{\mu}_\alpha \cos\delta, \overline{\mu}_\delta)= (0.37,-6.06)$~mas~yr$^{-1}$, with a mean error of 0.2~mas~yr$^{-1}$. 
We also show proper motions of five radio continuum sources in W40 derived by \cite{Ortiz2018}.
The maser source has a proper motion different from the motion of the \gaia sources located in the west of the Serpens South region. This is expected since the maser emission arises from gas that may be moving along a jet or along a disk wind \citep{Moscadelli2022} associated with CARMA--6. We note that the direction of the maser proper motion is close to perpendicular to the large-scale molecular outflow, which is oriented in the southeast-northwest direction (see Fig.~3 in \citealt{Ortiz2021AJ}). This suggests that the masers are most likely associated with a disk wind. We will investigate this in a forthcoming paper. 

On the other hand, the motions of most \gaia sources are larger in magnitude than the expected proper motion due to Galactic rotation 
(see, e.g., \citealt[][]{Dzib2017}) for objects at the location of the Serpens South and W40 
regions.
The expected proper motion is shown as a black arrow in Fig. \ref{fig:gaia-appendix} and has values of $\mu_\alpha \cos\delta =0.83$ and $\mu_\delta=-5.18$~mas~yr$^{-1}$. 

\begin{figure}[!bht]
\begin{center}
{\includegraphics[width=0.4\textwidth,angle=0]{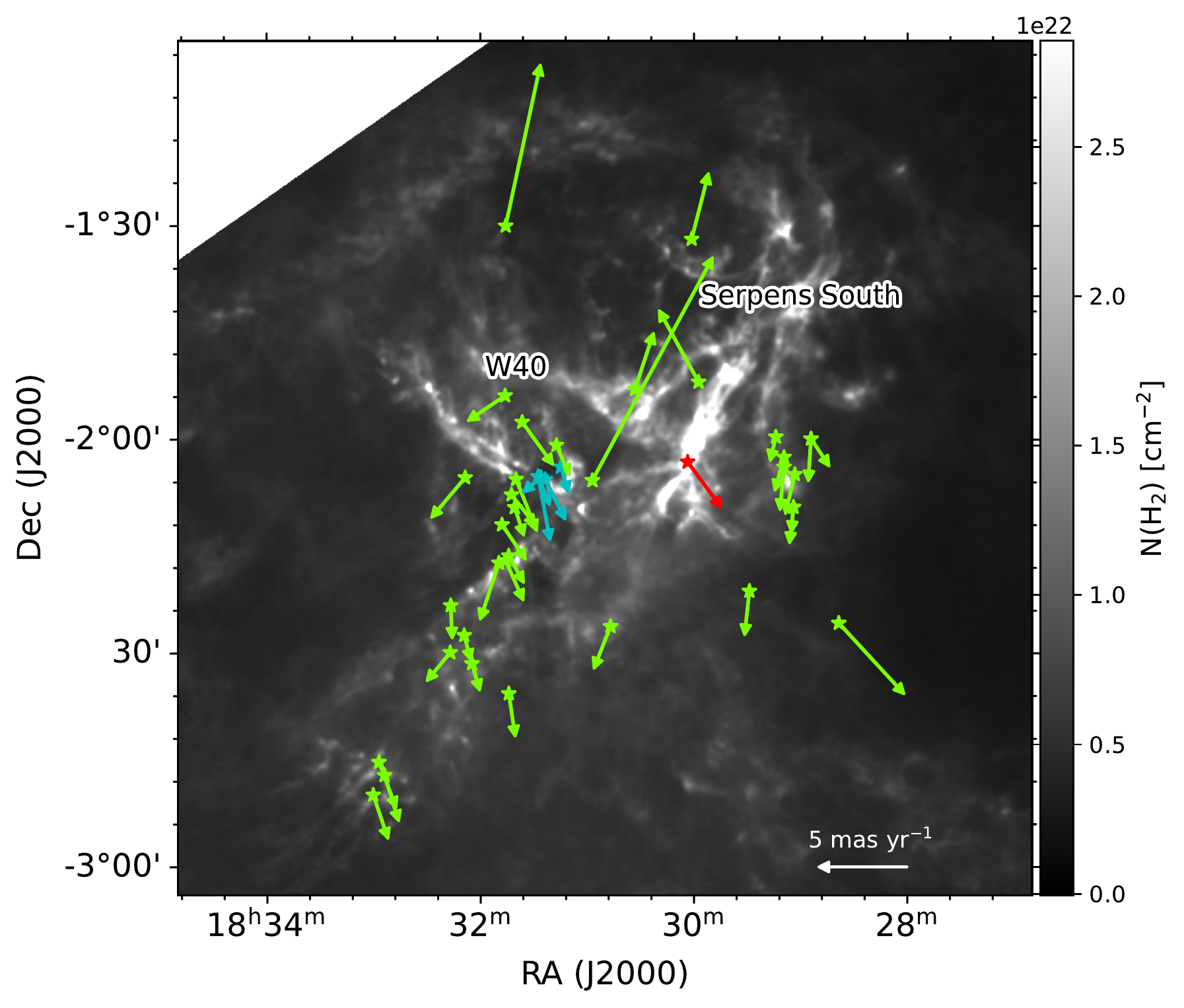}}
 \end{center}
\caption{Spatial distribution of \gaia sources (green star symbols) in the Serpens South and W40 
regions. The arrows indicate their proper motions. The red star and red arrow mark the location of the water maser associated with CARMA--6 and the proper motion from the astrometric fits. The cyan arrows indicate proper motions of stars in W40 that have VLBA astrometric solutions \citep{OrtizLeon2015}. We subtracted the expected proper motion of the Serpens South and W40 
regions due to Galactic rotation from the proper motions shown here. The background shows the H$_2$ column density map obtained with {\it Herschel}  \citep{Andre2010}. 
}
\label{fig:gaia}
\end{figure}

To obtain proper motions of sources in the rest frame of the molecular cloud complex, we had to remove the expected proper motion from Galactic rotation. We used a rotation velocity for the LSR of $\Theta_0=239$~\kms, a distance to the Galactic center of $R_0=8.3$~kpc \citep{Brunthaler2011}, and a Solar motion of $U_0=11.10$~\kms, $V_0=12.24$~\kms, and $W_0=7.25$~\kms ~\citep{Schnrich2010}.
The resulting stellar motions from this subtraction are shown in Fig. \ref{fig:gaia}. We note that in this figure sources in the west of Serpens South have proper motions mostly pointing toward the south, whereas sources in the east of W40 have RA\ proper motions mostly pointing toward the west. 
This may result from an interaction between W40 and Serpens South. The proper motions of stars in W40 have magnitudes $\approx 2$~mas~yr$^{-1}$, which correspond to tangential velocities $\approx 4$~\kms for a distance of 440.7~pc.

\cite{Shimoikura2019} suggest that an outer expanding shell created by the W40 HII region is interacting with the dense gas associated with the Serpens South young cluster, and that this interaction induced star formation in the cluster (see also \citealt{Nakamura2017}). \cite{Shimoikura2019} estimate an expansion velocity for the shell of $\sim$3~\kms. This is similar to the magnitude of the velocity of the \gaia sources in W40 ($\approx 4$~\kms). However, we see in Fig.~\ref{fig:gaia} that the \gaia sources at the westernmost side of Serpens South belong to ``filament A''  traced by C$^{18}$O at 6~\kms (see Fig.~12 of \citealt{Shimoikura2019}), and the model shown in their Fig.~18 suggests that this filament is not interacting with the outer boundary shell of the \object{W40 HII region}. In Appendix \ref{sec:gaia} all \gaia stars with $\varpi/\sigma_\varpi<5$ were excluded, a few of which were close to the Serpens South main filament. Thus, the origin of these motions could be investigated with higher precision astrometric data.

\section{Conclusions}\label{sec:conclusions}

We performed VLBA observations of 22~GHz masers arising from the protostar CARMA--6 in the Serpens South cluster. The high precision astrometry to maser spots allowed us to derive a distance of $440.7\pm$3.5~pc for this object. This protostar is at the heart of the cluster, whose young members are concentrated within molecular filaments \citep[e.g.,][]{Sun2022} that extend over an area of $\sim 20'\times10'$ on the sky ($\sim 2.6~{\rm pc}\times1.3$~pc at $d=440$ pc). The depth of the cluster could be $\sim$3~pc, assuming the main filament has a length along the line of sight similar to its length on the sky. We added in quadrature this value to the formal uncertainty and conservatively determine the distance to the young cluster to be $440.7\pm$4.6~pc. 

This result rules out shorter distances suggested in recent works (e.g., \citealt{winston2018} and \citealt{podio2021}). In addition, it is an improvement over the Serpens South distance obtained with \gaia DR3 sources in the least dense areas of the region and an improvement over previous estimates based on sources in the nearby regions W40 and Serpens Main. Furthermore, our results combined with previous VLBA parallaxes indicate that Serpens South, W40, and Serpens Main are all physically associated.


\begin{acknowledgements} 

G.N.O.L. acknowledges financial support from UNAM-DGAPA postdoctoral fellowship program. L.L.
acknowledges the support of DGAPA PAPIIT grant
IN112820 and CONACYT-CF grant 263356.

The National Radio Astronomy Observatory is a facility of the National Science Foundation operated under a cooperative agreement by Associated Universities, Inc.

This work has made use of data from the European Space Agency (ESA) mission \gaia ({\tt \url{https://www.cosmos.esa.int/gaia}}), processed by the \gaia Data Processing and Analysis Consortium (DPAC, {\tt \url{https://www.cosmos.esa.int/web/gaia/ dpac/consortium}}). Funding for the DPAC has been provided by national institutions, in particular the institutions participating in the \gaia Multilateral Agreement.

\end{acknowledgements}

\bibliographystyle{aa} 
\bibliography{ms.bib}

\begin{appendix}

\section{Supplementary figures and tables}\label{sec:supplm}

Table \ref{tab:obs-vlba} lists the details of the observations performed with the VLBA.
Table \ref{tab:jmfit} gives the positions of maser spots obtained from Gaussian fits to the brightness distribution and using the AIPS task {\tt{jmfit}}. 

\begin{figure*}[!htb]
\begin{center}
 {\includegraphics[width=0.8\textwidth,angle=0]{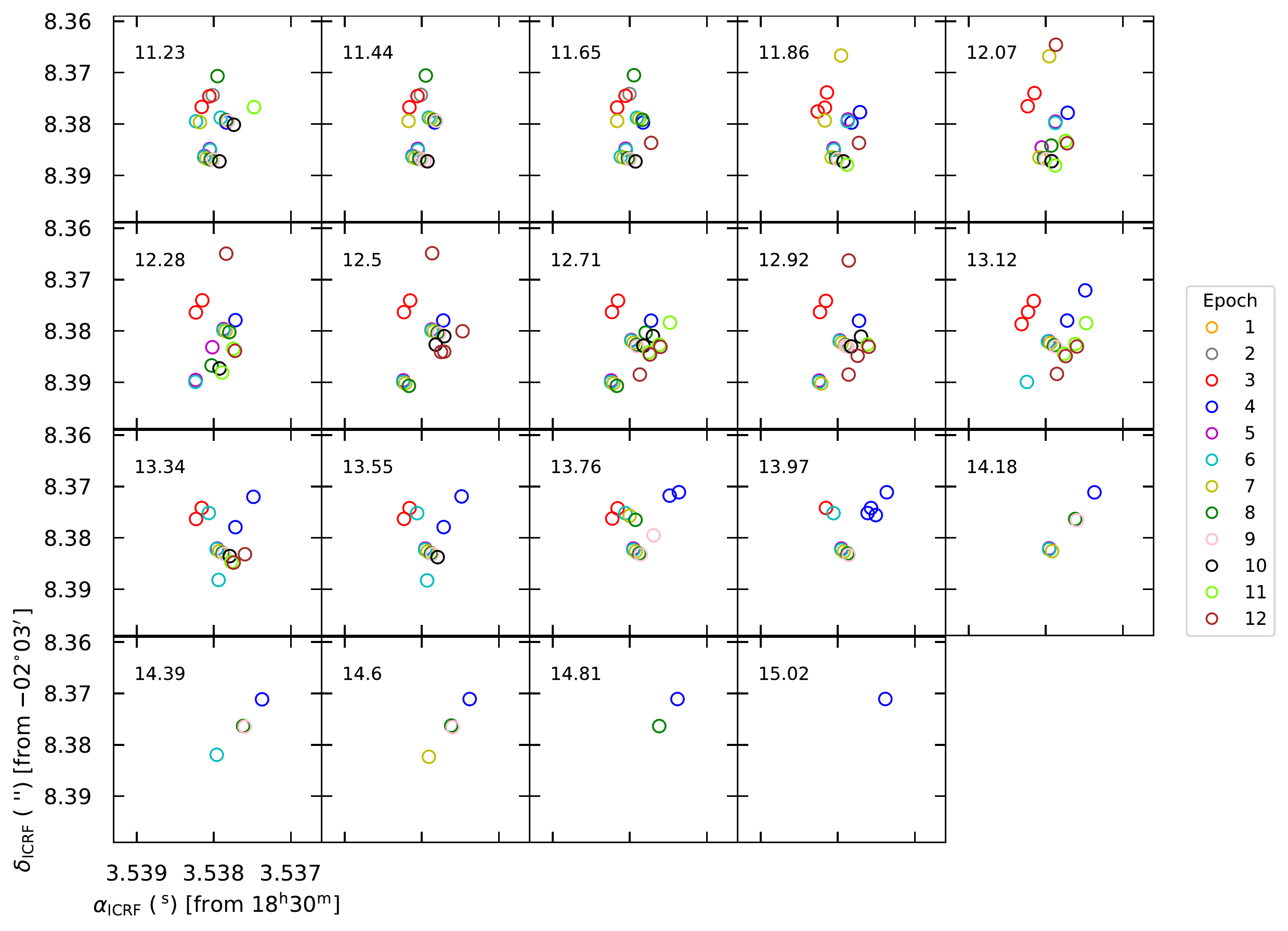}}
\caption{Similar to Fig. \ref{fig:spots} but for the velocity range 11.23 to 15.02 \kms.  
}
\label{fig:spots-2}
\end{center}
\end{figure*}

Figures \ref{fig:spots-2} and \ref{fig:spots-woprlx} show the distribution of the detected water maser spots with and without the parallax signature subtracted, respectively.
Figure \ref{fig:fit-combined-2} shows the astrometric fit to maser spots detected in epochs 4 to 9 and the velocity range 9.76--10.81~\kms.
Figures \ref{fig:fit-104}, \ref{fig:fit-106}, and \ref{fig:fit-1081} show the astrometric fits for velocity channels 10.4, 10.6, and 10.81~\kms, respectively.

\begin{table*}[!hb]
\tiny
\caption{VLBA observed epochs.}
\label{tab:obs-vlba} 
\centering 
\begin{tabular}{l l c c c c}  
\hline\hline  
\# & ID & Observation  & Beam      &  P.A.     &  Channel     \\
  &      & Date       & Size  &   & rms  \\
    &      &              & (mas$\times$mas) & ($^{\rm o}$)   & (mJy~beam$^{-1}$)  \\
\hline 
1 & BO061A3 & 2020 Sep  21 & 1.6$\times$0.3 & $-17$ &  9 \\
2 & BO061A4 & 2020 Oct  25 & 1.3$\times$0.4 & $-17$ &  8 \\
3 & BO061A5 & 2021 Apr 02 &  1.4$\times$0.4 & $-17$ &  9 \\
4 & BO061A6 & 2021 Sep 04 &  1.6$\times$0.3 & $-18$ & 10 \\
5 & BO068A &  2022 Mar 14 &  1.5$\times$0.4 & $-18$ &  7 \\
6 & BO068B & 2022 Apr 11 &   1.4$\times$0.6 & $-7$ &  10 \\ 
7 & BO068C & 2022 May 06 &   1.3$\times$0.4 & $-15$ & 10  \\
8 & BO068D & 2022 May 30 &   1.5$\times$0.4 & $-18$ &  9 \\ 
9 & BO068E & 2022 Jun 08 &   1.4$\times$0.3 & $-17$ & 12 \\
10 & BO068F & 2022 Jul 07 &  1.8$\times$0.3 & $-18$ & 13 \\ 
11 & BO068G & 2022 Aug 27 &  2.9$\times$0.3 & $-19$ & 15 \\
12 & BO068H & 2022 Sep 08 &  1.2$\times$0.3 & $-17$ & 12 \\ 
\hline 
\end{tabular}
\end{table*}

\begin{table*}[!hb]
\tiny
\caption{Maser spot positions.}
\label{tab:jmfit} 
\centering 
\begin{tabular}{c c c c c c c c c}  
\hline\hline  
 Velocity channel & Julian Day &  $\alpha$(J2000) & $\sigma_\alpha$ & $\delta$(J2000) & $\sigma_\delta$ &   \\
 (\kms) & & (h:m:s) & (s) & ($^{\rm o}$:$'$:$''$) &($''$)  \\
\hline 
9.76 & 2459114.56829 & 18:30:03.53822820 &  0.00000185 & -02:03:08.376282 & 0.000078 \\
& 2459148.47313 & 18:30:03.53818104 &  0.00000234 & -02:03:08.377222 & 0.000092 \\
& 2459307.04326 & 18:30:03.53832871 &  0.00000200 & -02:03:08.379177 & 0.000083 \\
& 2459462.61304 & 18:30:03.53791920 &  0.00000014 & -02:03:08.382839 & 0.000004 \\
& 2459462.61304 & 18:30:03.53788515 &  0.00000020 & -02:03:08.381776 & 0.000007 \\
& 2459653.08302 & 18:30:03.53813337 &  0.00000295 & -02:03:08.386172 & 0.000100 \\
& 2459705.93832 & 18:30:03.53809439 &  0.00000217 & -02:03:08.386643 & 0.000073 \\
\hline
9.97 & 2459114.56829 & 18:30:03.53822891 &  0.00000341 & -02:03:08.376468 & 0.000098 \\
& 2459307.04326 & 18:30:03.53833487 &  0.00000359 & -02:03:08.379005 & 0.000122 \\
& 2459462.61304 & 18:30:03.53790362 &  0.00000071 & -02:03:08.382417 & 0.000026 \\
& 2459462.61304 & 18:30:03.53791405 &  0.00000014 & -02:03:08.382718 & 0.000004 \\
& 2459653.08302 & 18:30:03.53813780 &  0.00000130 & -02:03:08.386316 & 0.000047 \\
& 2459681.00657 & 18:30:03.53813554 &  0.00000222 & -02:03:08.386480 & 0.000074 \\
\hline
10.18 & 2459462.61304 & 18:30:03.53788712 &  0.00000010 & -02:03:08.381886 & 0.000004 \\
& 2459653.08302 & 18:30:03.53813892 &  0.00000066 & -02:03:08.386292 & 0.000023 \\
& 2459681.00657 & 18:30:03.53813117 &  0.00000132 & -02:03:08.386483 & 0.000043 \\
& 2459705.93832 & 18:30:03.53809611 &  0.00000124 & -02:03:08.386650 & 0.000049 \\
\hline
10.40 & 2459307.04326 & 18:30:03.53822214 &  0.00000171 & -02:03:08.380405 & 0.000067 \\
& 2459462.61304 & 18:30:03.53789370 &  0.00000053 & -02:03:08.382123 & 0.000022 \\
& 2459653.08302 & 18:30:03.53813846 &  0.00000039 & -02:03:08.386357 & 0.000014 \\
& 2459681.00657 & 18:30:03.53813106 &  0.00000050 & -02:03:08.386486 & 0.000015 \\
& 2459705.93832 & 18:30:03.53809904 &  0.00000055 & -02:03:08.386761 & 0.000019 \\
& 2459729.87277 & 18:30:03.53804777 &  0.00000303 & -02:03:08.386955 & 0.000088 \\
\hline
10.60 & 2459307.04326 & 18:30:03.53822791 &  0.00000132 & -02:03:08.380376 & 0.000050 \\
& 2459462.61304 & 18:30:03.53788743 &  0.00000008 & -02:03:08.381954 & 0.000003 \\
& 2459653.08302 & 18:30:03.53813450 &  0.00000034 & -02:03:08.386322 & 0.000011 \\
& 2459681.00657 & 18:30:03.53813058 &  0.00000026 & -02:03:08.386534 & 0.000009 \\
& 2459705.93832 & 18:30:03.53809796 &  0.00000026 & -02:03:08.386766 & 0.000009 \\
& 2459729.87277 & 18:30:03.53804620 &  0.00000087 & -02:03:08.387044 & 0.000028 \\
\hline
10.81 & 2459307.04326 & 18:30:03.53823056 &  0.00000110 & -02:03:08.380358 & 0.000041 \\
& 2459462.61304 & 18:30:03.53788776 &  0.00000009 & -02:03:08.381989 & 0.000003 \\
& 2459653.08302 & 18:30:03.53812683 &  0.00000025 & -02:03:08.386320 & 0.000008 \\
& 2459681.00657 & 18:30:03.53812948 &  0.00000021 & -02:03:08.386523 & 0.000007 \\
& 2459705.93832 & 18:30:03.53809741 &  0.00000023 & -02:03:08.386759 & 0.000008 \\
& 2459729.87277 & 18:30:03.53804580 &  0.00000041 & -02:03:08.387017 & 0.000013 \\
& 2459738.84548 & 18:30:03.53800918 &  0.00000296 & -02:03:08.386988 & 0.000079 \\
\hline 
\end{tabular}
\end{table*}


\begin{figure*}[!bht]
\begin{center}
 {\includegraphics[width=0.8\textwidth,angle=0]{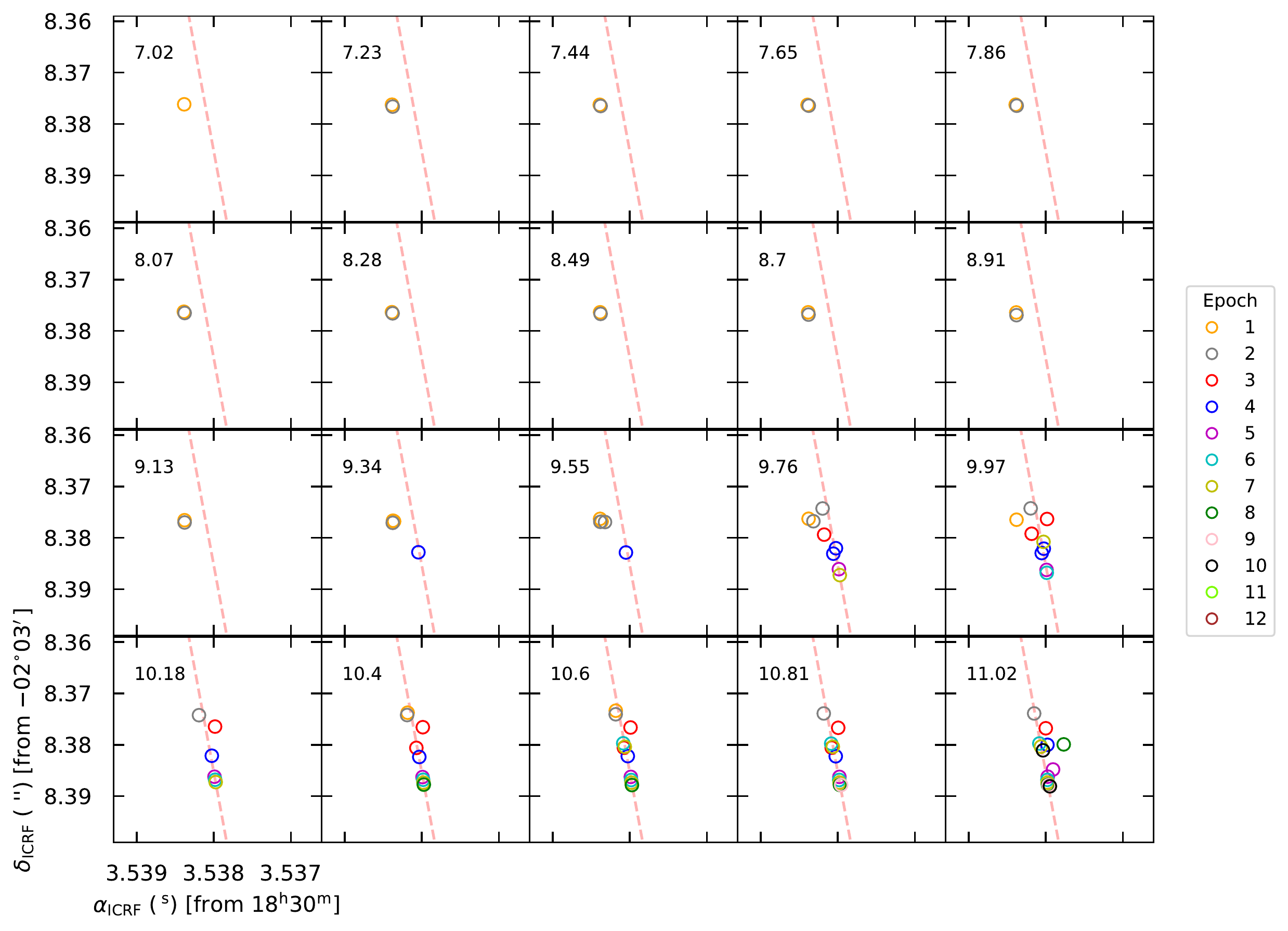}}
\caption{
Spatial distribution of the water maser spots. The parallax signature has been subtracted from all positions presented here. The numbers in the corners indicate the velocity channel in~\kms. 
The symbols are color-coded by epoch (see the figure legend).
}
\label{fig:spots-woprlx}
\end{center}
\end{figure*}

\setcounter{figure}{1}
\renewcommand{\thefigure}{A.2}

\begin{figure*}[!htb]
\begin{center}
 {\includegraphics[width=0.8\textwidth,angle=0]{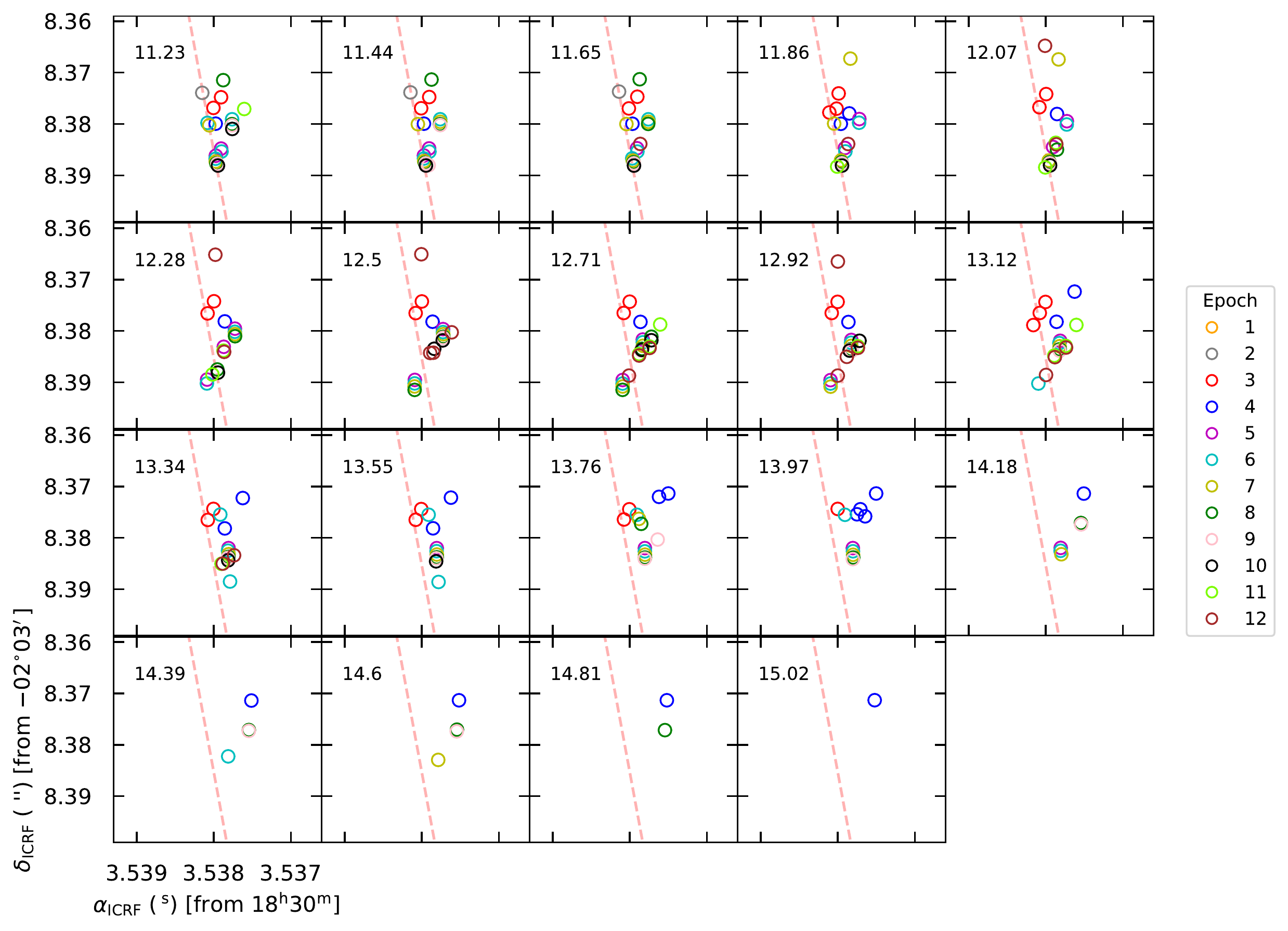}}
\caption{continued.}
\label{fig:spots-2-woprlx}
\end{center}
\end{figure*}

\setcounter{figure}{2}
\renewcommand{\thefigure}{A.\arabic{figure}}

\begin{figure*}[tbh]
\begin{subfigure}[t]{0.24\textwidth}
\centering
 \includegraphics[scale=0.34,angle=0]{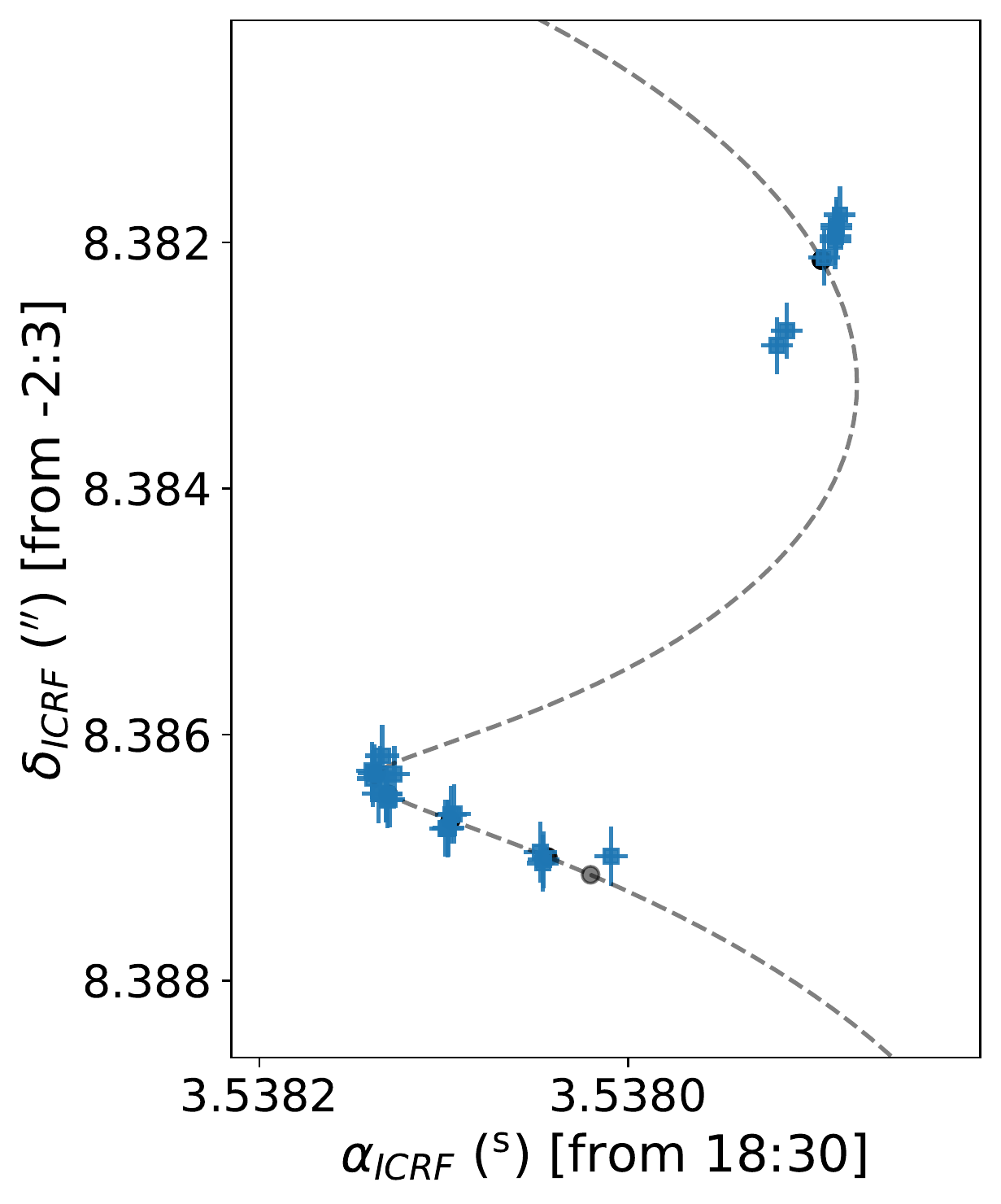}
 \caption{}
 \end{subfigure}
 \begin{subfigure}[t]{0.36\textwidth}
 \centering
 \includegraphics[scale=0.32,angle=0]{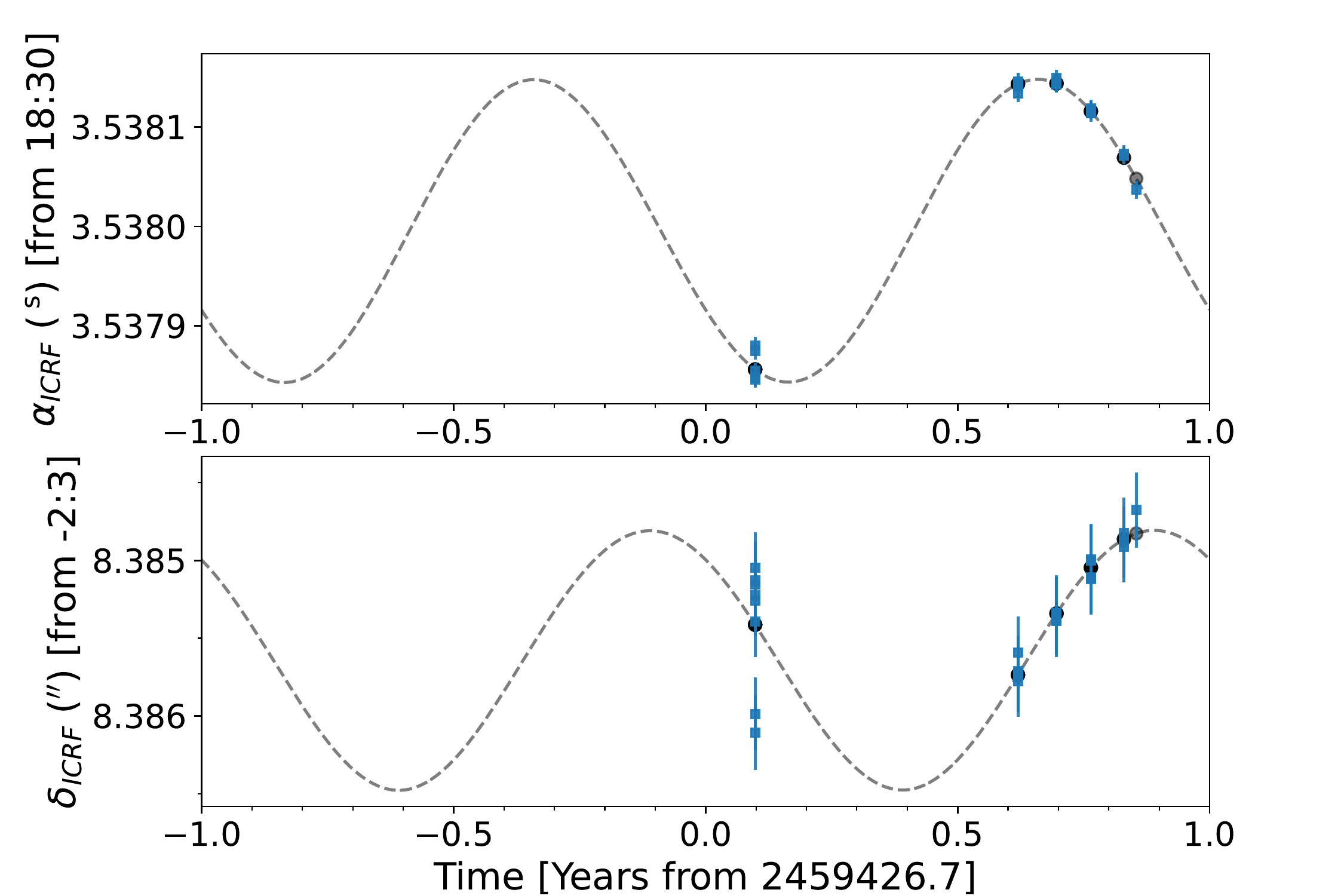}
  \caption{}
  \end{subfigure} 
  \begin{subfigure}[t]{0.3\textwidth} 
 \includegraphics[scale=0.32,angle=0]{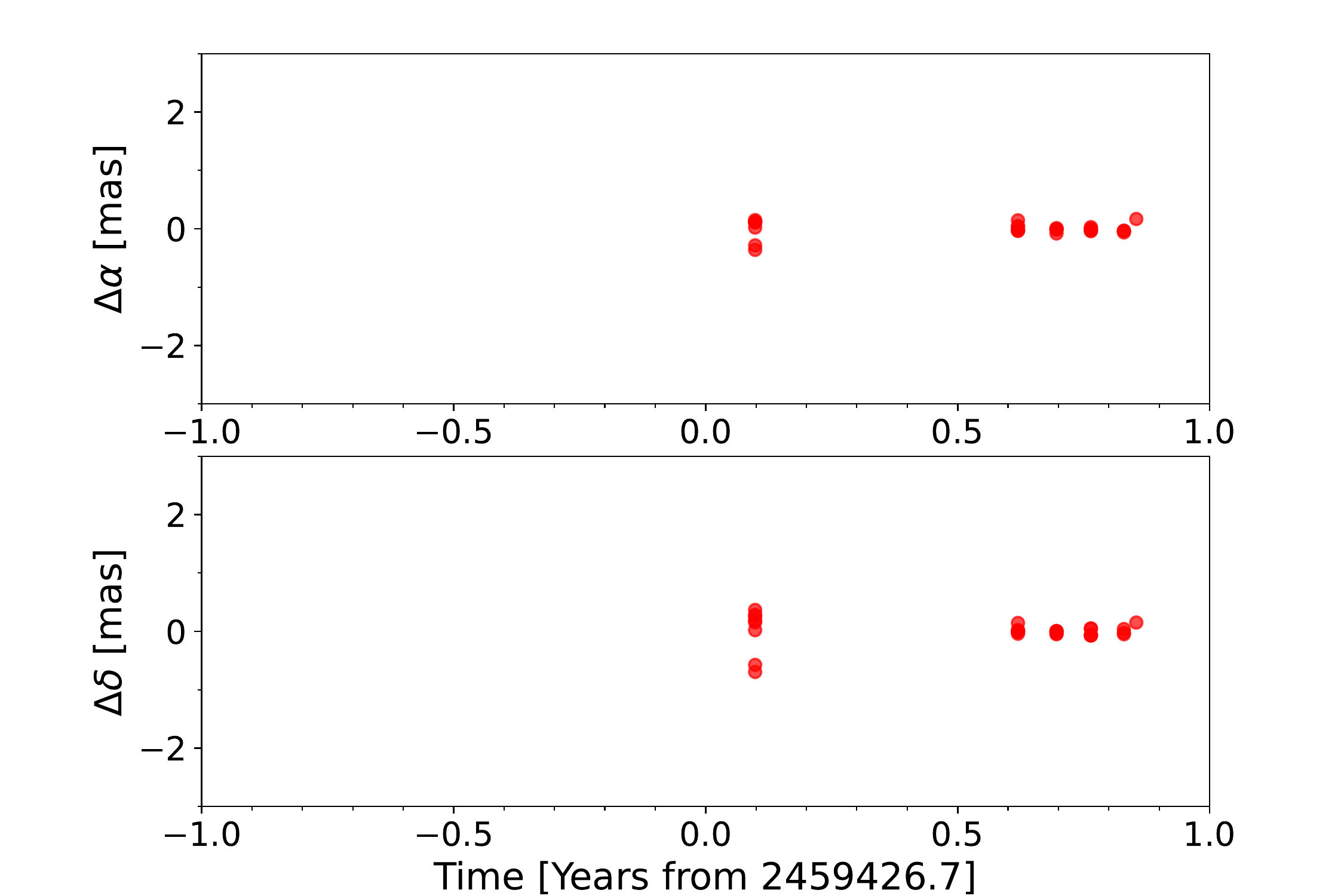}
  \caption{}
   \end{subfigure} 
\caption{Same as Fig. \ref{fig:fit-combined} but after removing spots from epochs 1 to 3. }
\label{fig:fit-combined-2}
\end{figure*}

\begin{figure*}[tbh]
\begin{subfigure}[t]{0.24\textwidth}
\centering
 \includegraphics[scale=0.34,angle=0]{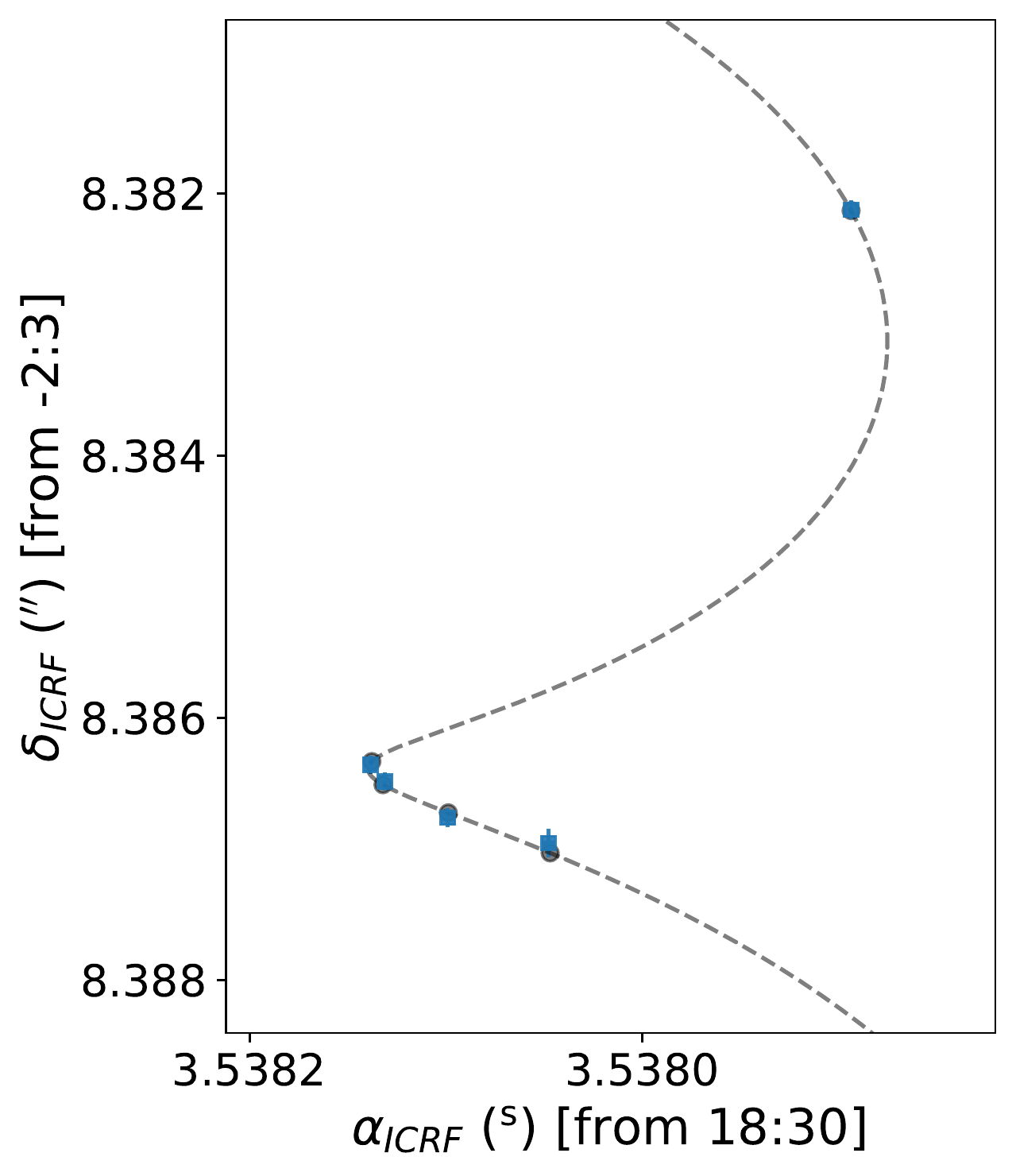}
 \caption{}
 \end{subfigure}
 \begin{subfigure}[t]{0.36\textwidth}
 \centering
 \includegraphics[scale=0.32,angle=0]{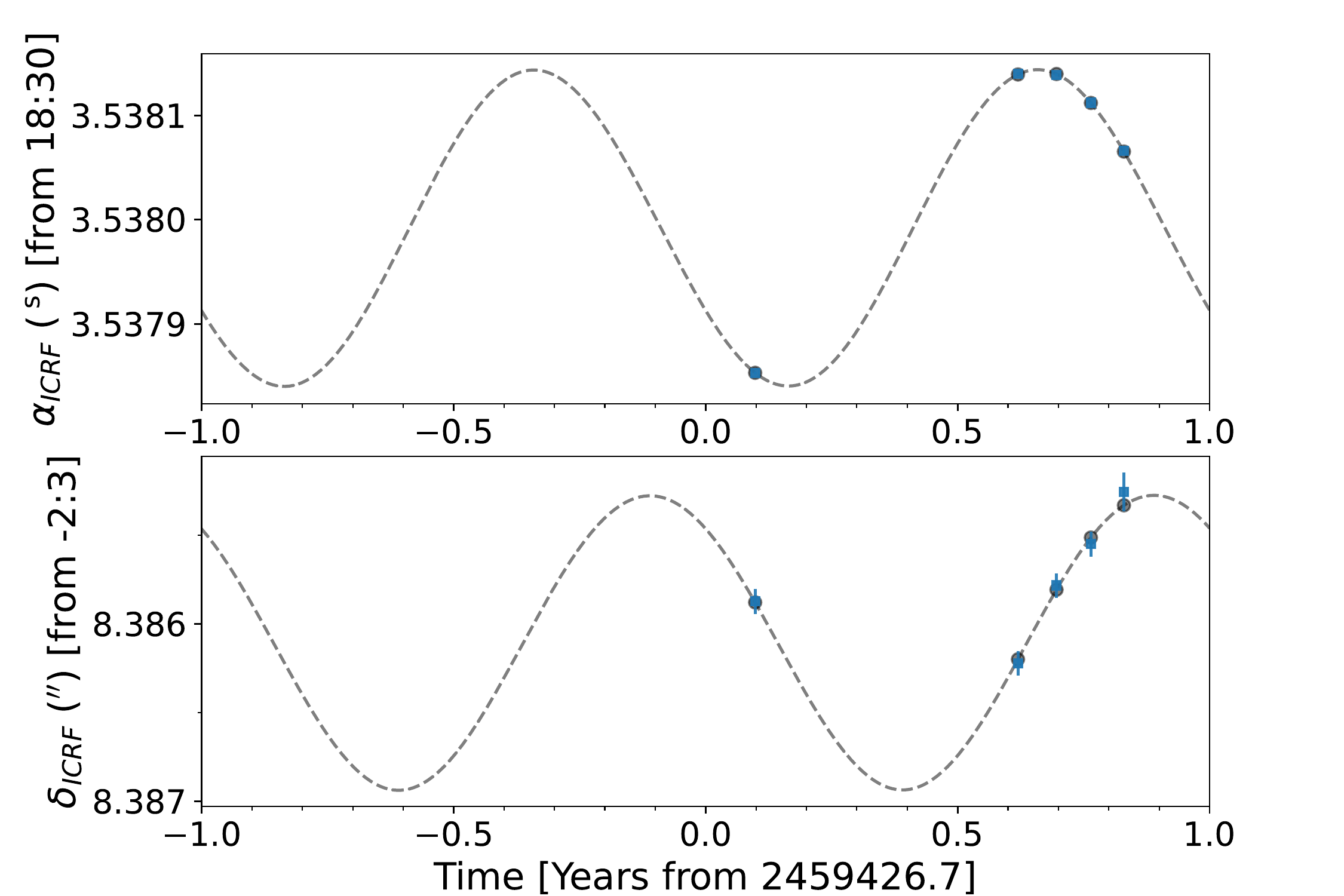}
  \caption{}
  \end{subfigure} 
  \begin{subfigure}[t]{0.3\textwidth} 
 \includegraphics[scale=0.32,angle=0]{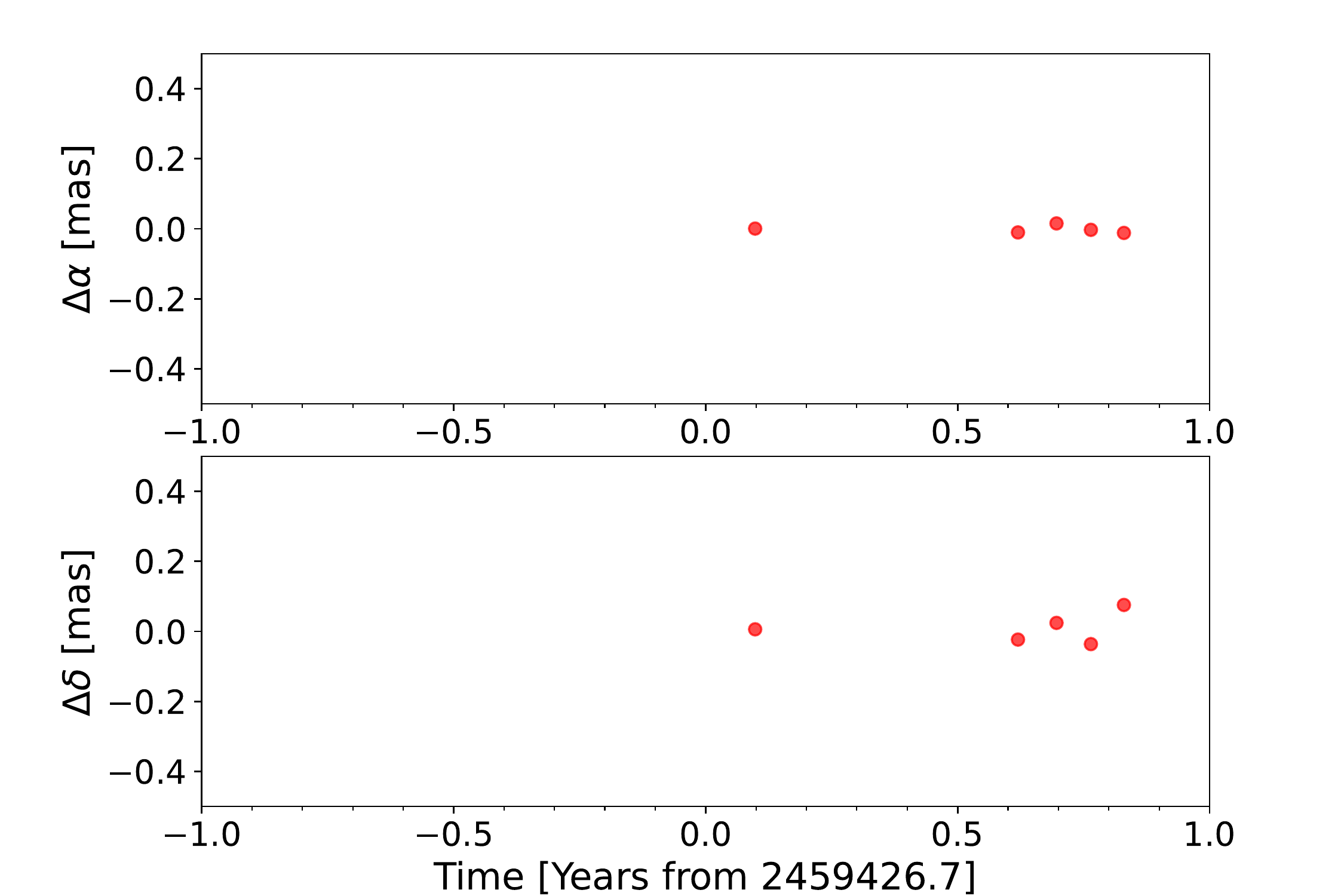}
  \caption{}
   \end{subfigure} 
\caption{Same as Fig. \ref{fig:fit-combined-2} but for the 10.40~\kms ~channel.  }
\label{fig:fit-104}
\end{figure*}

\begin{figure*}[tbh]
\begin{subfigure}[t]{0.24\textwidth}
\centering
 \includegraphics[scale=0.34,angle=0]{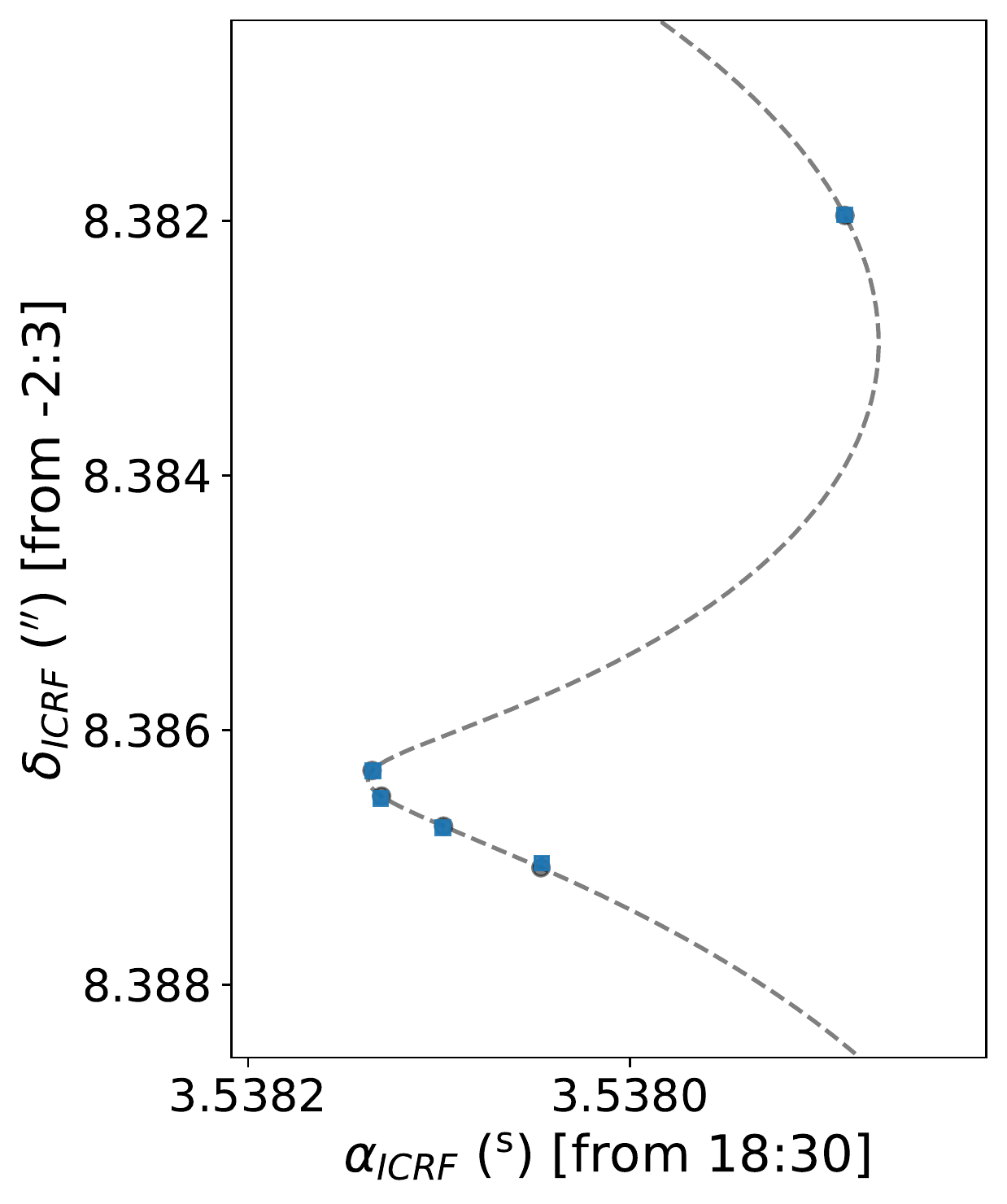}
 \caption{}
 \end{subfigure}
 \begin{subfigure}[t]{0.36\textwidth}
 \centering
 \includegraphics[scale=0.32,angle=0]{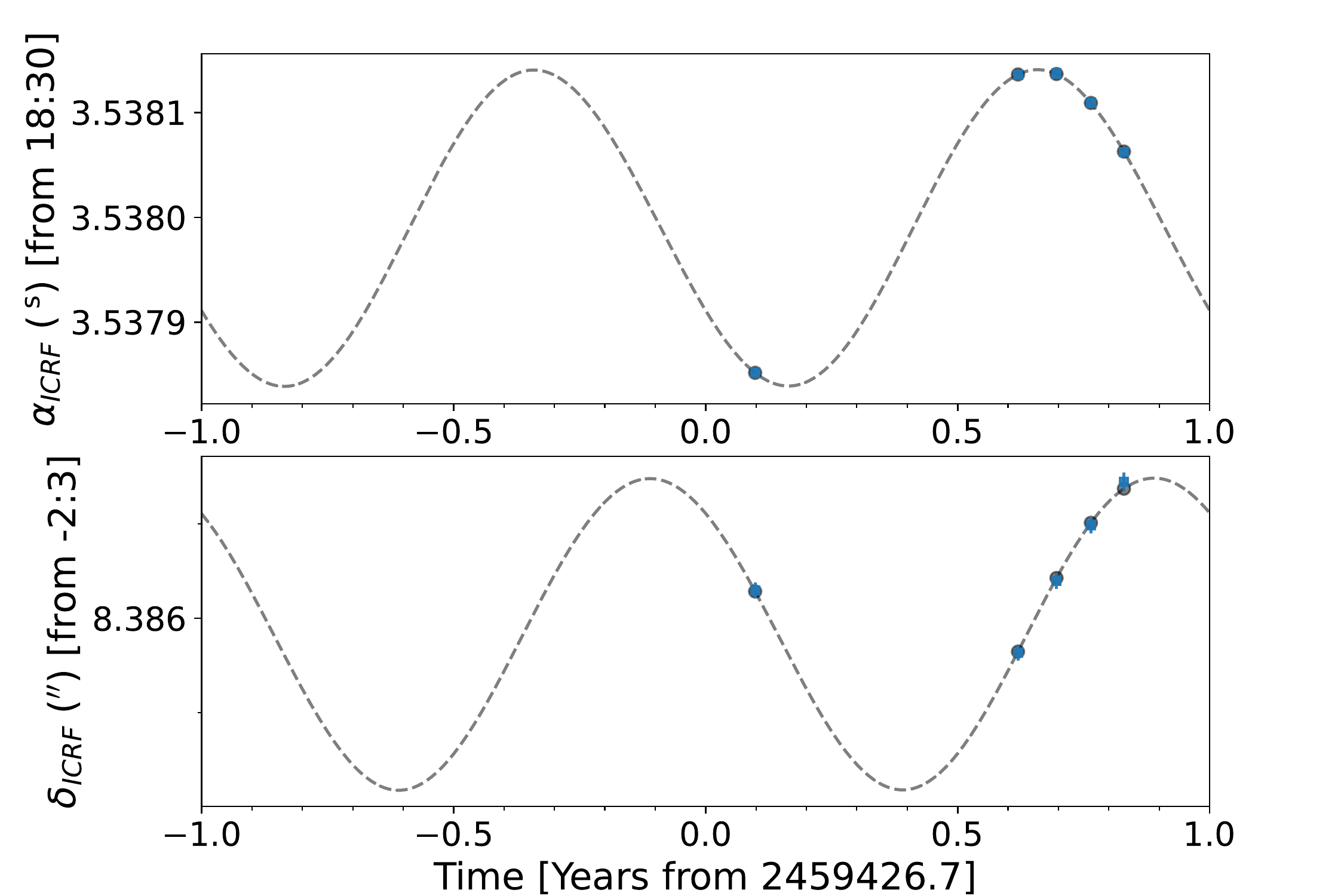}
  \caption{}
  \end{subfigure} 
  \begin{subfigure}[t]{0.3\textwidth} 
 \includegraphics[scale=0.32,angle=0]{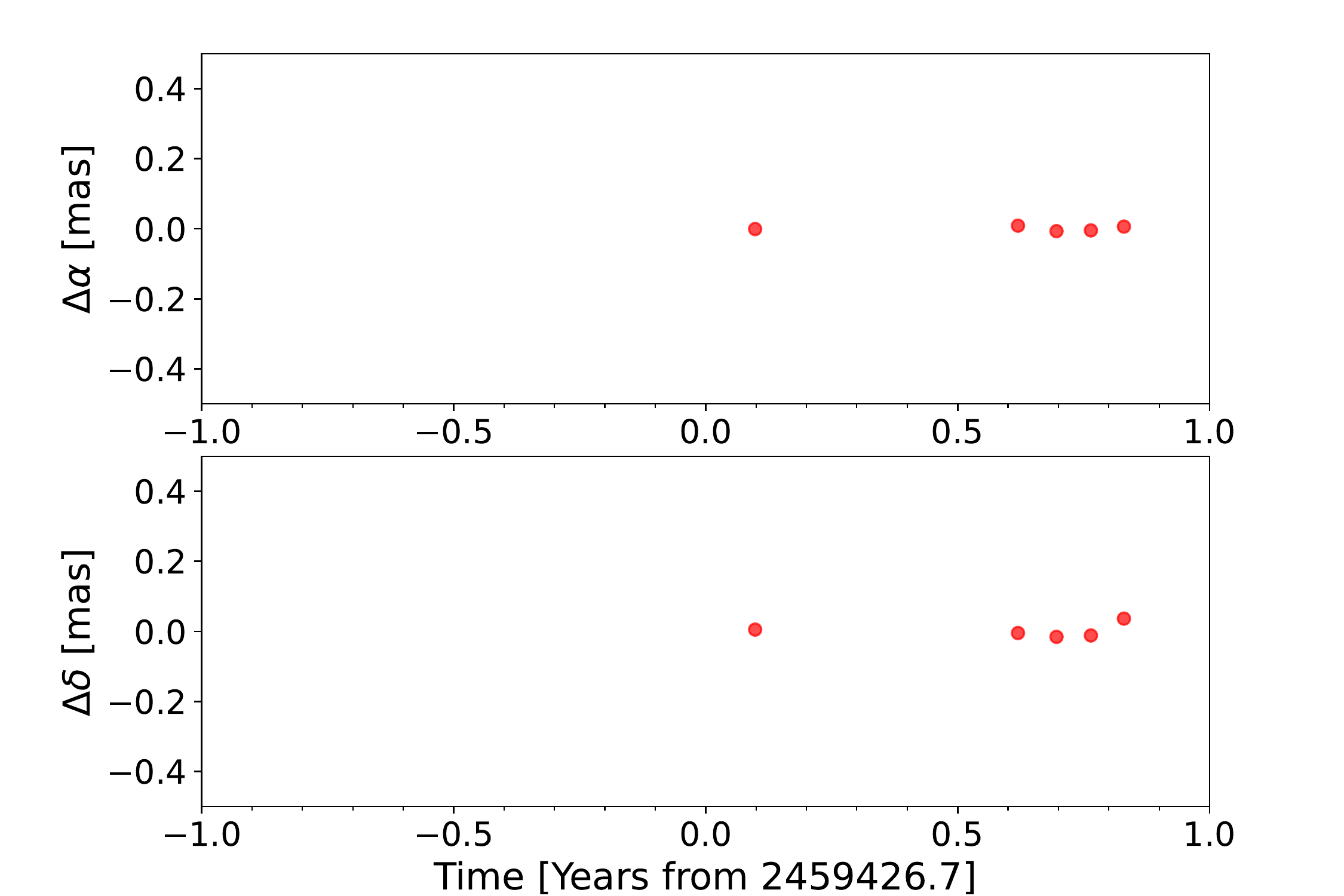}
  \caption{}
   \end{subfigure} 
\caption{Same as Fig. \ref{fig:fit-combined-2} but for the 10.60~\kms ~channel.  }
\label{fig:fit-106}
\end{figure*}

\begin{figure*}[tbh]
\begin{subfigure}[t]{0.24\textwidth}
\centering
 \includegraphics[scale=0.34,angle=0]{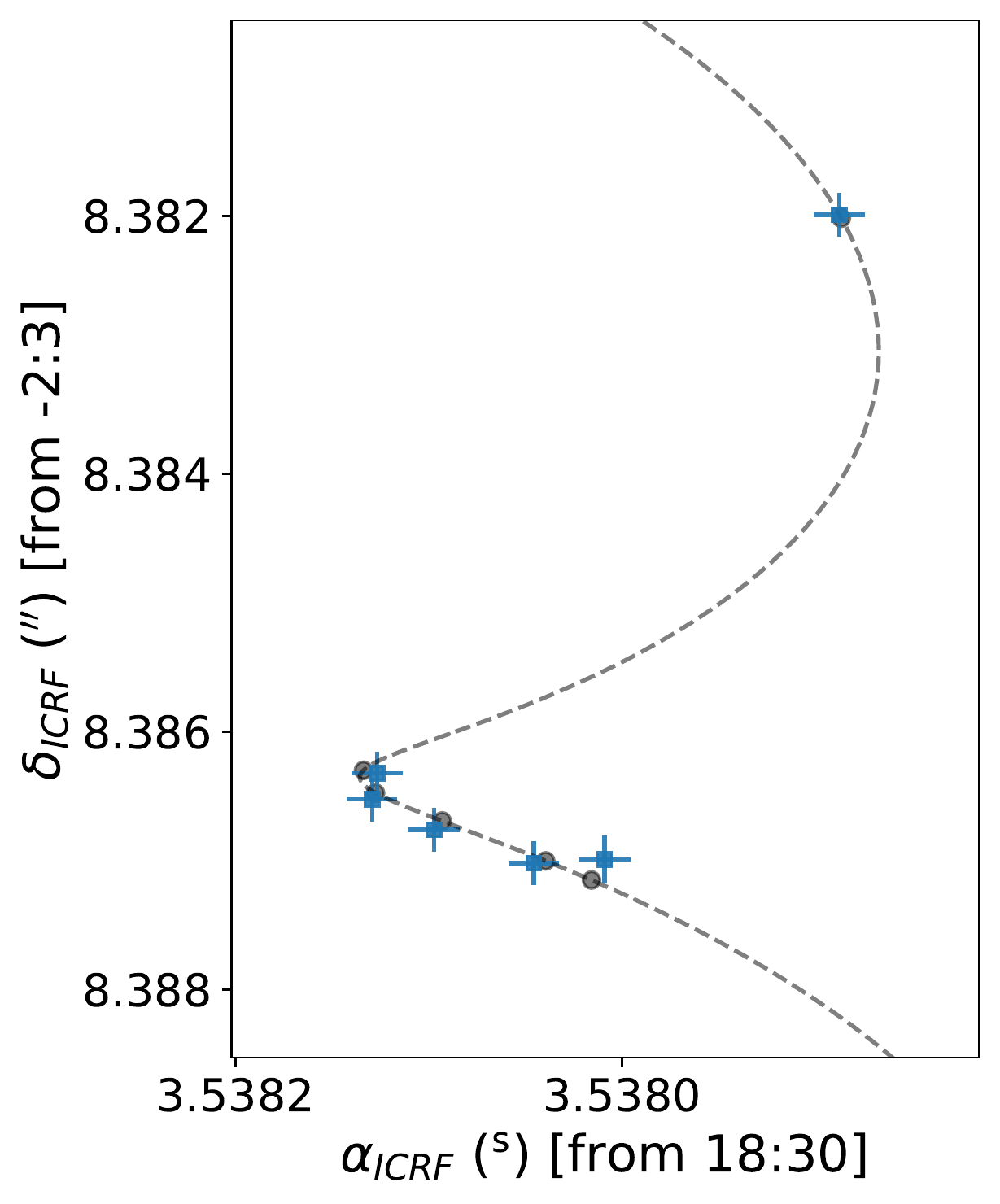}
 \caption{}
 \end{subfigure}
 \begin{subfigure}[t]{0.36\textwidth}
 \centering
 \includegraphics[scale=0.32,angle=0]{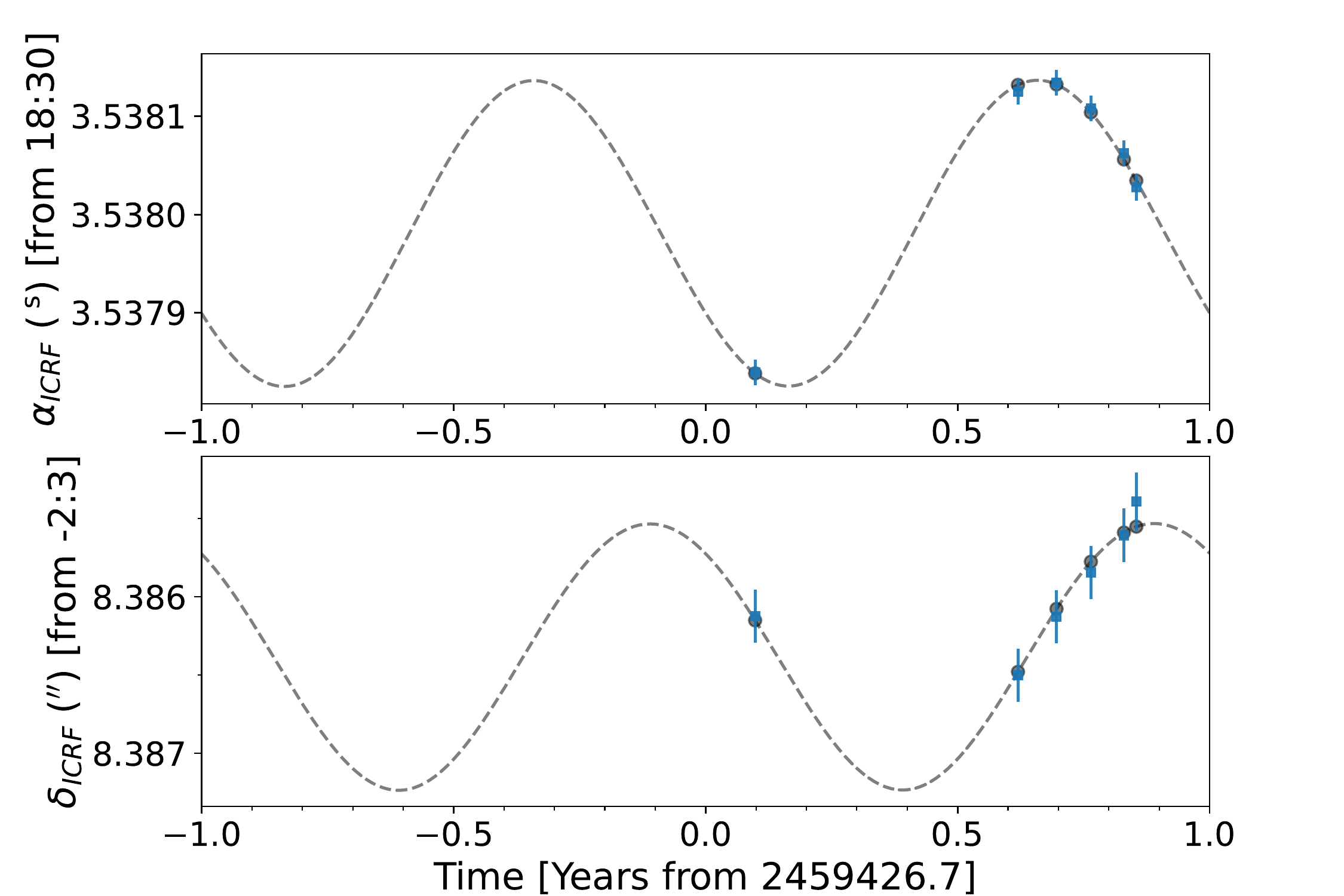}
  \caption{}
  \end{subfigure} 
  \begin{subfigure}[t]{0.3\textwidth} 
 \includegraphics[scale=0.32,angle=0]{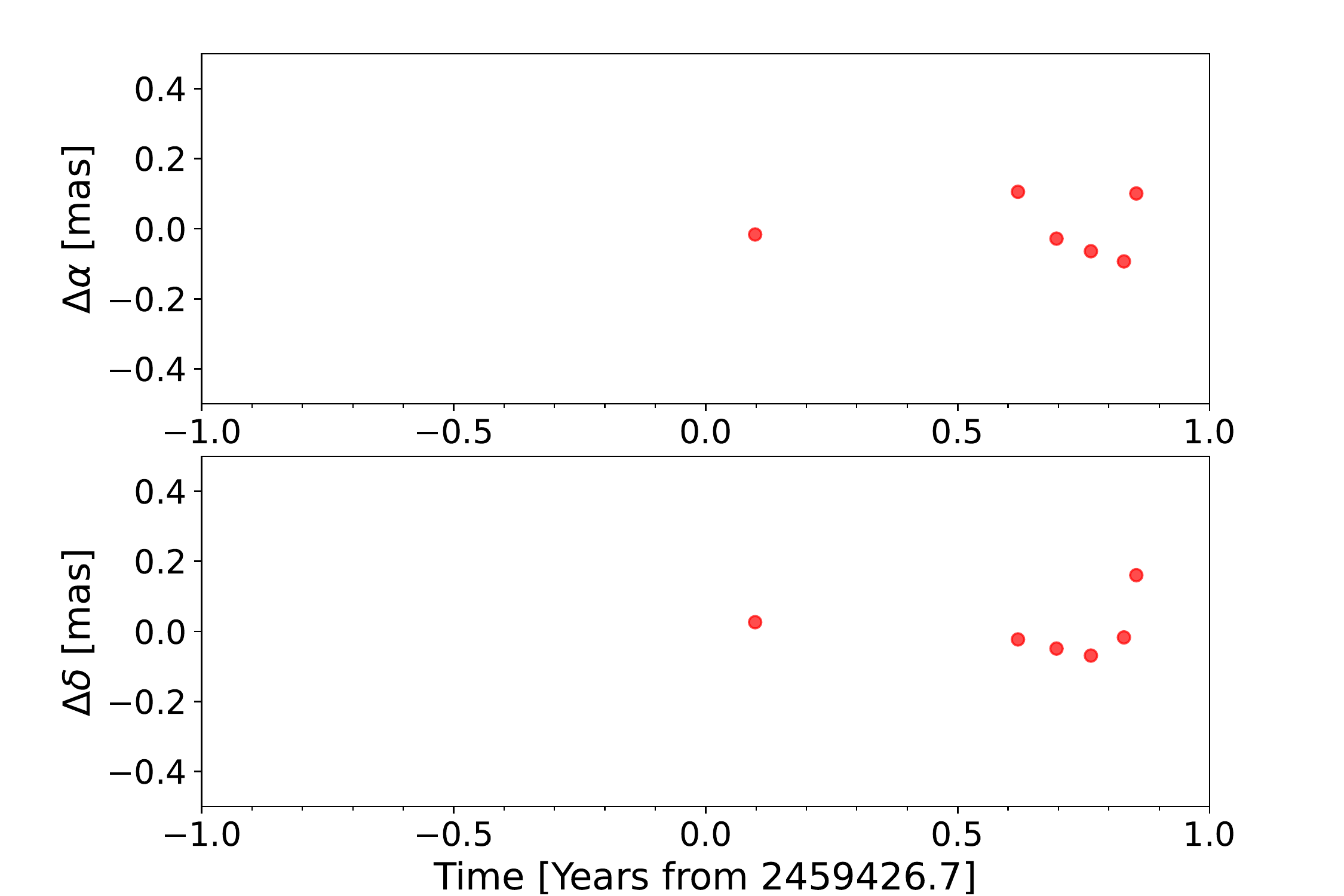}
  \caption{}
   \end{subfigure} 
\caption{Same as Fig. \ref{fig:fit-combined-2} but for the 10.81~\kms ~channel.  }
\label{fig:fit-1081}
\end{figure*}

\clearpage

\section{Astrometric fits taking an unseen companion into account }\label{sec:companion}

It is possible that the large residuals seen in Fig.~\ref{fig:fit-combined} are due to an unresolved companion that is associated with CARMA--6 and perturbs its motion. To test this possibility, specifically of an unseen companion in a long-period orbit 
($\gg$1~yr), we fit the maser data with a model that includes acceleration terms. The results are presented in Fig.~\ref{fig:fit-acceleration}. The parallax and distance resulting from this fit are $2.16\pm0.10$~mas and $464\pm22$~pc, respectively. The rms of the residuals (0.34~mas in both RA\ and Decl.) are smaller than the residuals from the fit that does not include acceleration by an unseen companion (Fig.~\ref{fig:fit-combined} and Table \ref{tab:fits}). However, except for the first two epochs, we see that the residuals 
show a similar temporal trend
as those shown Fig.~\ref{fig:fit-combined}. When including acceleration terms, the data from the third and fourth epochs still show large deviations from the best fit (see Fig. \ref{fig:fit-acceleration}). Thus, a model that takes an unseen companion in a long-period orbit into account does not produce a better fit to the full data set (9.76--10.81 \kms\ velocity range and nine epochs). 

We also investigated the possibility that there is a short-period companion perturbing the motion of CARMA--6. Following \cite{Curiel2020}, we fit a model that includes a Keplerian orbit induced on the star by an unseen companion. We explored initial orbital periods from 0.2 to 2~yr but did not find a reasonable fit to the data, as neither the period nor eccentricity converge.

\setcounter{figure}{0}
\renewcommand{\thefigure}{B.\arabic{figure}}

\begin{figure}[!hbt]
\begin{subfigure}[b]{0.49\textwidth}
\centering
 \includegraphics[scale=0.38,angle=0]{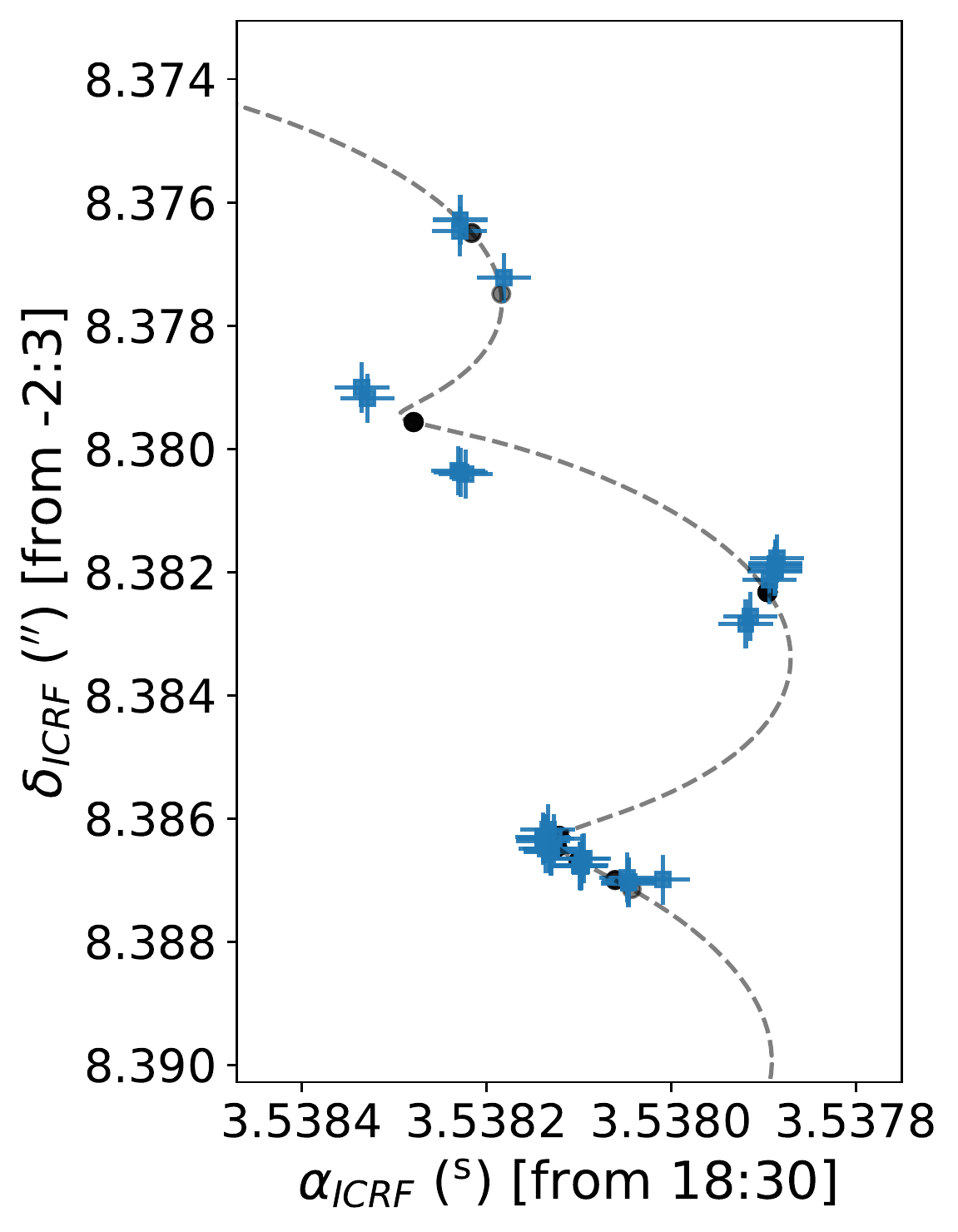} 
 \caption{}
 \end{subfigure} 
\hfill
 \begin{subfigure}[b]{0.49\textwidth}
 \centering
 \includegraphics[scale=0.34,angle=0]{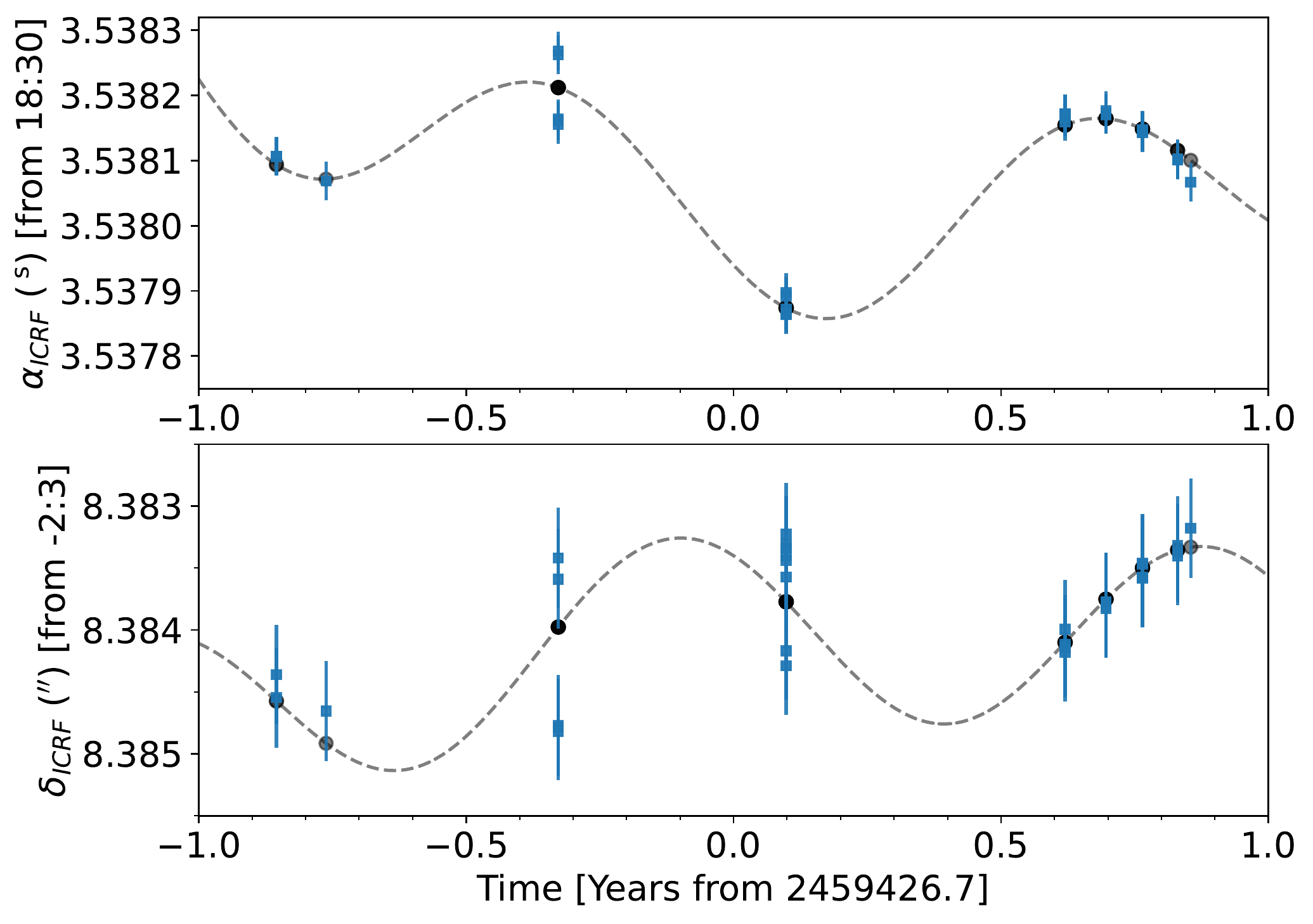}
  \caption{}
  \end{subfigure}
  \hfill
  \begin{subfigure}[b]{0.49\textwidth} 
 \centering
 \includegraphics[scale=0.34,angle=0]{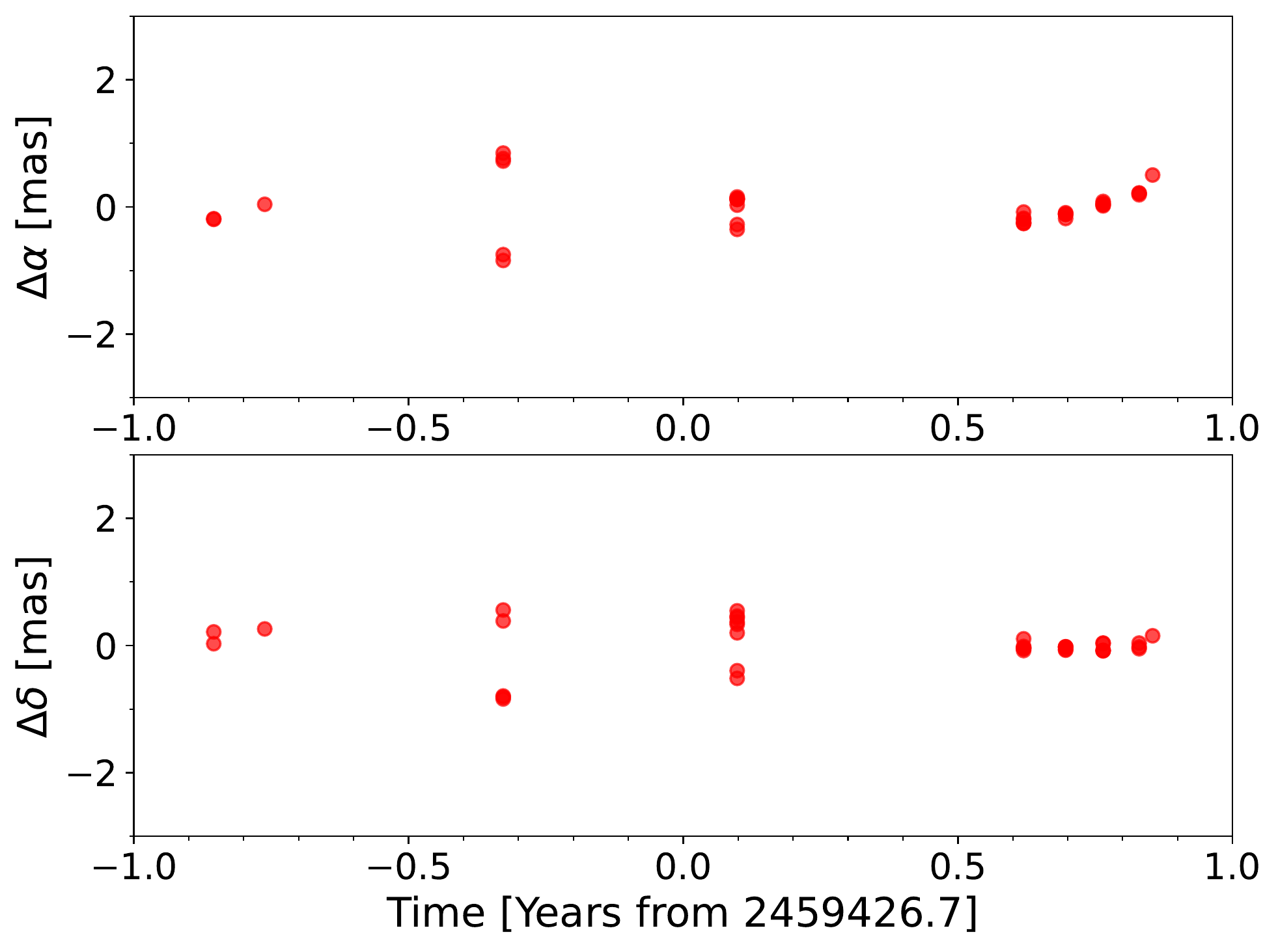}
  \caption{}
   \end{subfigure} 
   \hfill
\caption{Same as Fig. \ref{fig:fit-combined} but including acceleration terms in the fitted model.  }
\label{fig:fit-acceleration}
\end{figure}

\clearpage

\section{Comparison with \gaia DR3}\label{sec:gaia}

We followed a similar analysis to that done in \cite{Ortiz2018} in order to compare the distance obtained from the water masers with that derived from \gaia DR3 (\citealt{Gaia2022}). We used the catalog of YSO candidates from the {\it Spitzer} Core two Disks (c2d) and Gould Belt survey \citep[][]{Dunham_2015}.  This catalog contains 1319 sources in the Aquila Rift, including Serpens South and W40. We cross-matched the positions of the YSO candidates against \gaia DR3 positions using a matching radius of 1$''$ and the Virtual Observatory tool TOPCAT \citep{Taylor_2005}. Then, we selected all sources with the following criteria: (i) $18^{\rm h}28^{\rm m}00^{\rm s}\leq\alpha\leq18^{\rm h}34^{\rm m}00^{\rm s}$,
(ii) $-3^{\rm o}\leq\delta\leq-1^{\rm o}$, (iii) $\varpi/\sigma_\varpi \geq 5$, and (iv) renormalized unit weight error (RUWE) $\leq1.4$ \citep{lindegren_ruwe}.\ This resulted in 21 sources. 

We also used the list of 316 Class I to Class III YSOs of \cite{Anderson2022}, and found that 46 of them are in the DR3 catalog and satisfy the criteria listed above. Combining them with the c2d sample by matching their \gaia IDs resulted in 35 sources. The distribution of these parallaxes is shown in Fig. \ref{fig:hist-gaia}, while Figs. \ref{fig:gaia} and \ref{fig:gaia-appendix} show the spatial distribution and proper motions of the 35 sources. We see that all of them are projected in the least dense parts of the Serpens South and W40 regions, while the maser source is located at the center of the Serpens South cluster. 
 The parallaxes show a Gaussian distribution (Fig. \ref{fig:hist-gaia}), with a peak and standard deviation of $2.11\pm0.02$~mas and 0.38~mas, respectively. 
 We note, however, that there are a few sources outside the distribution, with parallaxes above 3~mas and below 1 mas, respectively. 

Excluding parallaxes $\varpi>3$~mas and $\varpi<1$~mas, we find that the weighted mean parallax of the \gaia stars is 
$\varpi = 2.29\pm 0.23$~mas, 
which translates to 
$d = 437^{+49}_{-40}$~pc. 
This distance is consistent within $1\sigma$ with the water maser distance.

We also used the \gaia DR3 catalog to select all sources with (i) $r\leq15'$, (ii) $1 \leq \varpi [{\rm mas}] \leq 5$, (iii) $\varpi/\sigma_\varpi \geq 5$, and (iv) RUWE $\leq1.4$, where $r$ is the angular separation from the center of the Serpens South region (at $18^{\rm h}30^{\rm m}00^{\rm s}$ --$02^{\rm o}00^{\rm m}00^{\rm s}$). The parallax distribution of these stars is shown in Fig. \ref{fig:hist-gaia-2}. As expected,  when the sample is not restricted to known or candidate YSOs, the resulting parallax distribution is broader than the one shown in the histogram of Fig. \ref{fig:hist-gaia}. To estimate the width and peak of this distribution, we fit a Gaussian function. The peak is at $2.61\pm0.04$~mas, and the width is 0.7~mas. The parallax peak corresponds to a distance of $d=382.5$~pc. The weighted mean of the distribution is $2.84\pm0.71$~mas ($d=353^{+118}_{-71}$~pc), where the error bar is the standard deviation. 

This broader distribution with a larger parallax peak results from the contribution of stars that do not belong to the Serpens South cluster.
Therefore, larger parallaxes (or smaller distances) in this sample would be associated with a group of stars located at the front edge of the Aquila Rift complex but that are not related to the young cluster, which is embedded in the Serpens South region.

The value found above of $d=353$~pc is close to the \textit{Gaia}-based distance used by \cite{galametz2019} and \cite{podio2021}. We speculate that \cite{galametz2019} identified a jump in extinction around 300–350 pc and assigned the distance to this extinction layer as the distance to the Serpens South cloud (the authors do not provide details for their distance derivation). However, the analysis presented above shows that the YSOs are embedded in a more distant layer.

\setcounter{figure}{0}
\renewcommand{\thefigure}{C.\arabic{figure}}

\begin{figure}[!bht]
\begin{center}
{\includegraphics[width=0.4\textwidth,angle=0]{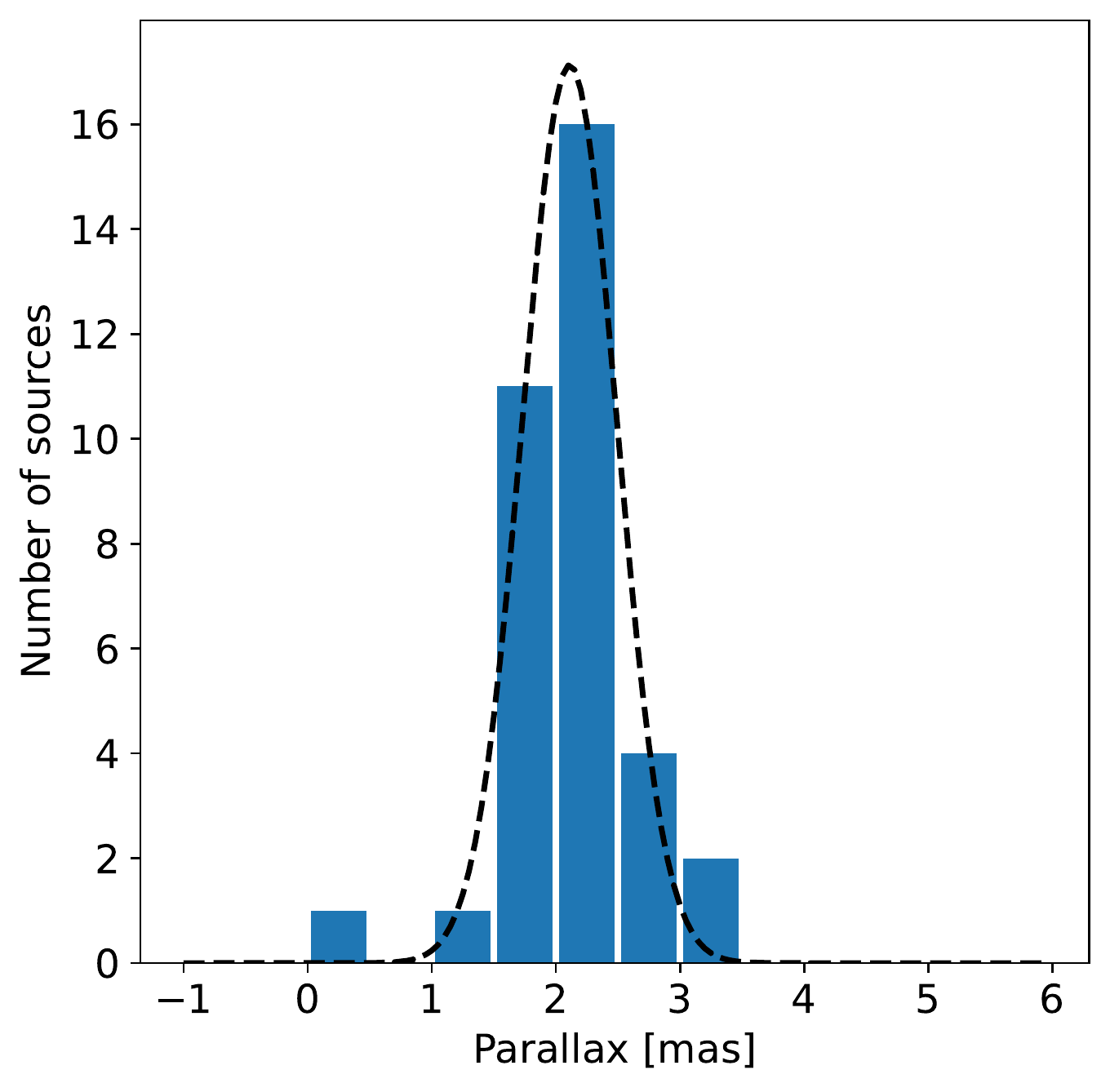}}
 \end{center}
\caption{Distribution of \gaia DR3 parallaxes for the sample of YSOs selected from the catalogs published by \cite{Dunham_2015} and \cite{Anderson2022} and that satisfy the criteria listed in Sect.~\ref{sec:gaia}. The dashed black line is a Gaussian fit to the distribution.
}
\label{fig:hist-gaia}
\end{figure}

\begin{figure}[!bht]
\begin{center}
{\includegraphics[width=0.4\textwidth,angle=0]{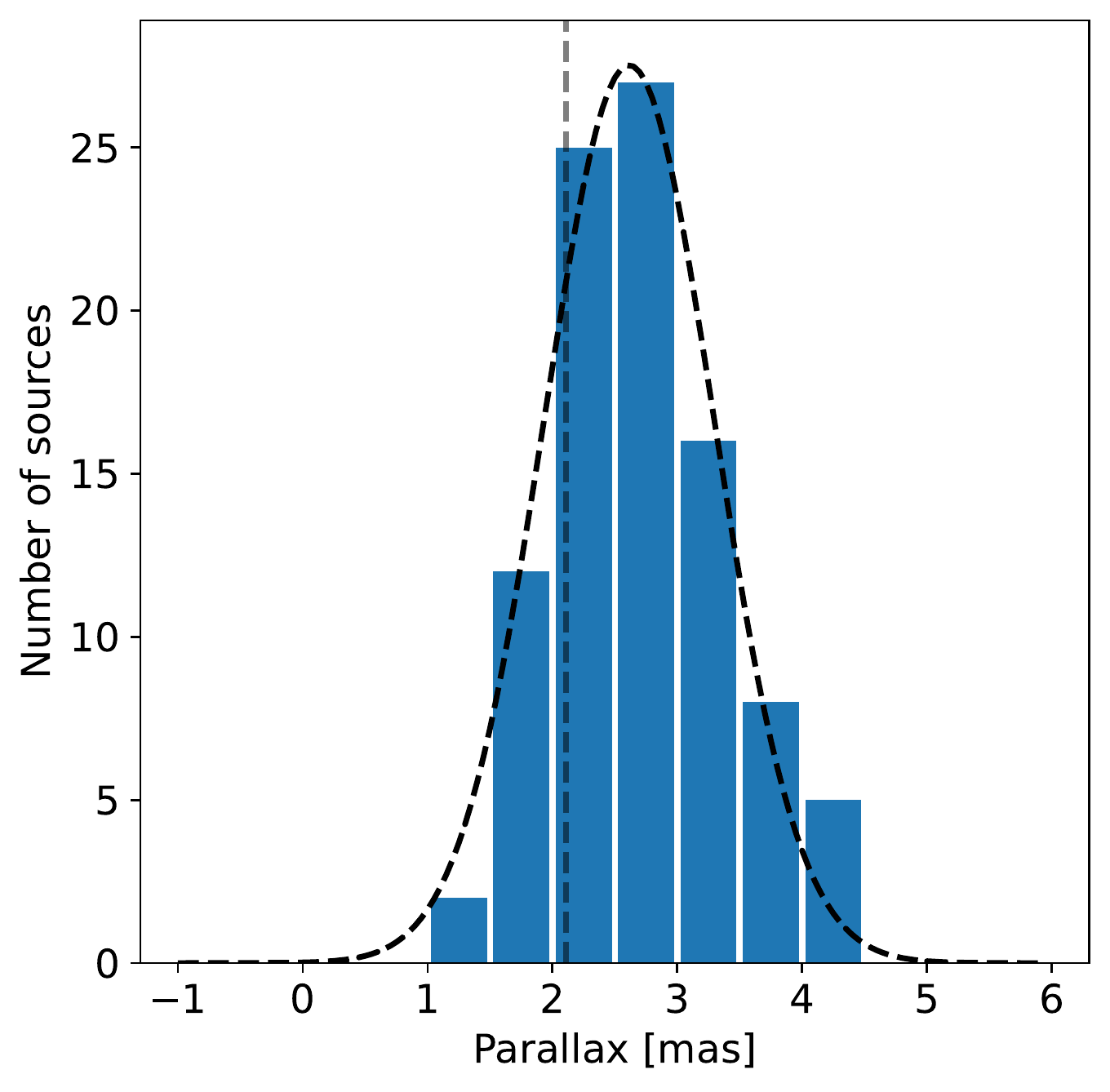}}
 \end{center}
\caption{Distribution of \gaia DR3 parallaxes for stars within $15'$ of the Serpens South center and that satisfy the criteria listed in Sect.~\ref{sec:gaia}. The vertical line indicates the peak of the parallax distribution shown in Fig.~\ref{fig:hist-gaia}, which only includes sources classified as YSOs ($\varpi=2.11$~mas). The dashed black line is a Gaussian fit to estimate the width and peak of the distribution.    
}
\label{fig:hist-gaia-2}
\end{figure}

\begin{figure}[!bht]
\begin{center}
{\includegraphics[width=0.49\textwidth,angle=0]{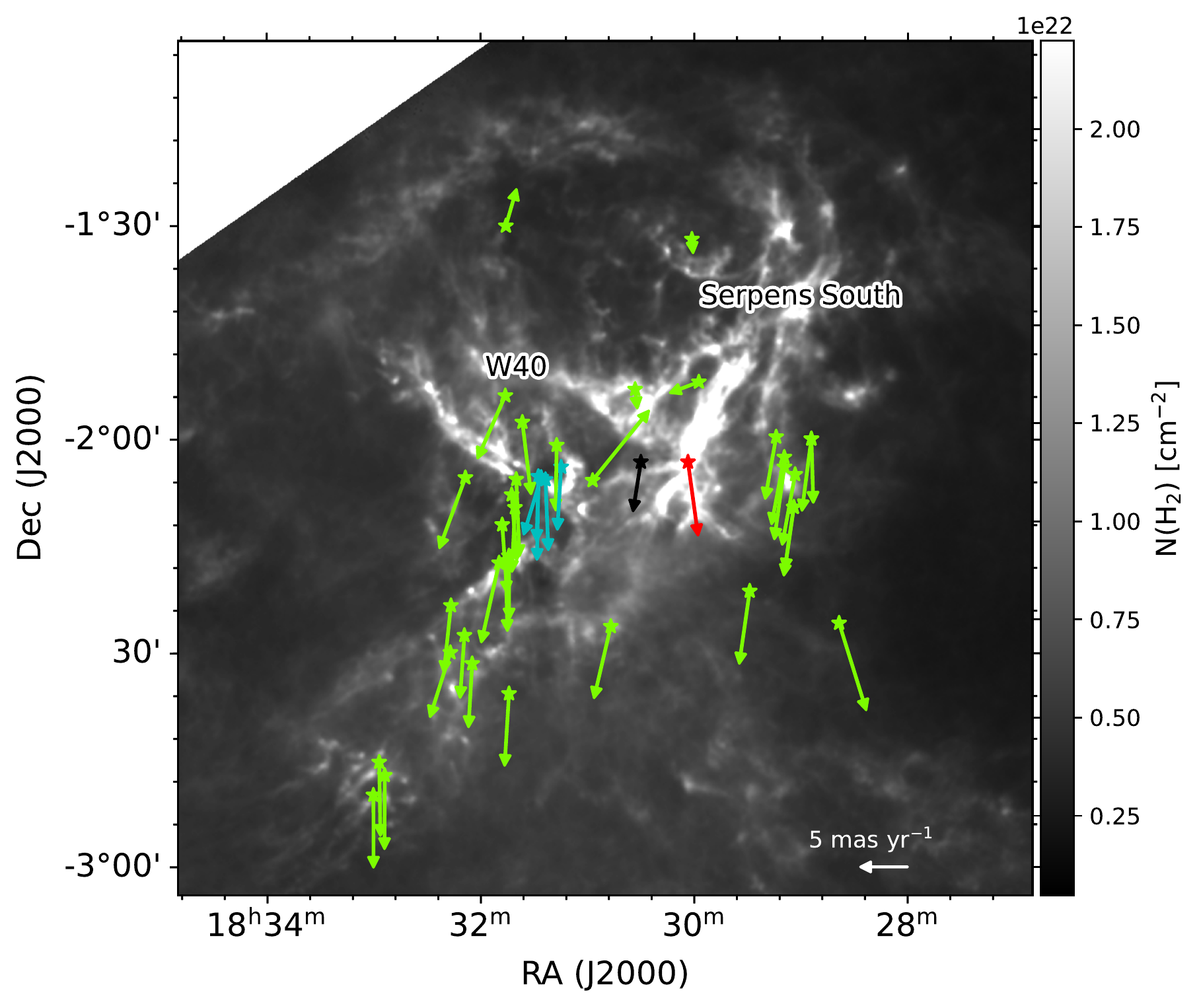}}
 \end{center}
\caption{Spatial distribution of \gaia sources (green star symbols) in the Serpens South/W40 region. The arrows indicate their published proper motions. The red star and red arrow mark the location of the water maser associated with CARMA--6 and the proper motion from the astrometric fits. The cyan arrows indicate proper motions of stars in W40 that have VLBA astrometric solutions \citep{OrtizLeon2015}. The black arrow indicates the expected proper motion of Serpens South/W40 from Galactic rotation.
The background shows the H$_2$ column density map obtained with {\it Herschel}  \citep{Andre2010}. }
\label{fig:gaia-appendix}
\end{figure}

\end{appendix}


\end{document}